\documentclass[12pt,a4paper]{article}
 \usepackage{amssymb,booktabs,subfig,xcolor,todonotes,microtype,setspace,bm,dsfont}
\usepackage[inline]{enumitem}
\usepackage[top=1.3in, bottom=1.3in, left=1.1in, right=1.1in]{geometry}
\usepackage{natbib}
\usepackage{url} % not crucial - just used below for the URL
\usepackage[pdftex,colorlinks=true,hypertexnames=false]{hyperref}
\definecolor{darkblue}{rgb}{0,0,.6}
\hypersetup{citecolor=darkblue,linkcolor=darkblue,urlcolor=darkblue}
\usepackage{amsmath}
\usepackage{amsfonts}
\usepackage{epsfig}
\usepackage{graphics}
\usepackage{palatino,eulervm}
\usepackage{graphicx}
\usepackage{lscape}
\usepackage{soul}
\usepackage{rotating,orcidlink}
\usepackage[linewidth=1pt]{mdframed}
\allowdisplaybreaks[4]
\DeclareMathOperator*{\argmin}{arg\,min}

\providecommand{\U}[1]{\protect\rule{.1in}{.1in}}
\newcommand{\X}{\mathcal{X}}
\newcommand{\Y}{\mathcal{Y}}
\newcommand{\Z}{\mathcal{Z}}
\newcommand{\E}{\text{E}}

\graphicspath{{plots/}}

\usepackage{amsthm,thmtools}
\usepackage{mathrsfs}

\makeatletter
\def\th@newremark{\th@remark\thm@headfont{\bfseries}}
\makeatletter
\theoremstyle{newremark}

\declaretheoremstyle[
  spaceabove=6pt, spacebelow=6pt,
  headfont=\bfseries,
  notefont=\mdseries, notebraces={(}{)},
bodyfont=\normalfont,
  postheadspace=0.5em,
]{mystyle}

\newcommand{\Rlogo}{\protect\includegraphics[height=1.8ex,keepaspectratio]{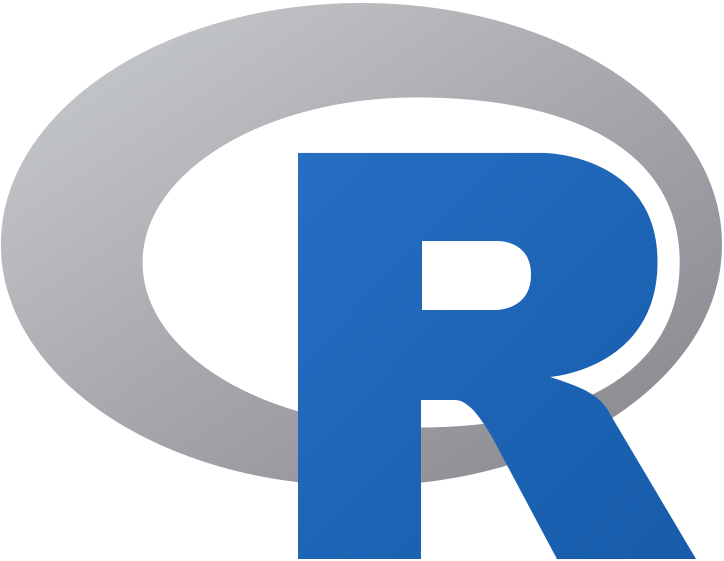}}

\begin{document}

\title{Nonstationary functional time series forecasting}
\author{{\hspace{-.3in} \normalsize Han Lin Shang\orcidlink{0000-0003-1769-6430}} \hspace{1.05in} {\normalsize Yang Yang\orcidlink{0000-0002-8323-1490}\thanks{
 Postal address: SR114, School of Information and Physical Sciences, The University of Newcastle, Newcastle, NSW 2308, Australia; Telephone: +61(2) 4921 8622; Email: yang.yang10@newcastle.edu.au}}
\\
{\normalsize Macquarie University\hspace{.56in} University of Newcastle}
 }

\date{\normalsize This version: \today}

\maketitle

\centerline{\bf Abstract}

\medskip

We propose a nonstationary functional time series forecasting method with an application to age-specific mortality rates observed over the years. The method begins by taking the first-order differencing and estimates its long-run covariance function. Through eigen-decomposition, we obtain a set of estimated functional principal components and their associated scores for the differenced series. These components allow us to reconstruct the original functional data and compute the residuals. To model the temporal patterns in the residuals, we again perform dynamic functional principal component analysis and extract its estimated principal components and the associated scores for the residuals. As a byproduct, we introduce a geometrically decaying weighted approach to assign higher weights to the most recent data than those from the distant past. Using the Swedish age-specific mortality rates from 1751 to 2022, we demonstrate that the weighted dynamic functional factor model can produce more accurate point and interval forecasts, particularly for male series exhibiting higher volatility.

\medskip


\setstretch{1.5}

\newpage

\section{Introduction}\label{sec:1}
\renewcommand{\theequation}{1.\arabic{equation}}
\setcounter{equation}{0}

In many developed countries, increases in human longevity due to decreasing mortality have led to government concerns regarding the sustainability of pensions, healthcare, and aged-care systems. The growing attention towards mortality modelling and forecasting for the benefit of making adequate government policies has arisen from these concerns. One of the policies relying on accurate mortality forecasts involves the determination of the qualifying age for a sustainable pension system \citep[see, e.g.,][for the case of Australia]{HZS21, CSY24}. For instance, Australians born after 2010 enjoy a life expectancy of over 80 years, which is around a 40\% increase since the start of the 20\textsuperscript{th} century \citep{AIHW22}. The extended longevity brings additional risks to the Australian superannuation system as more people would have insufficient savings to support them through their retirement without recourse to the Age Pension \citep{Sup22}. Also, age-specific mortality forecasts are important inputs to the calculation of remaining life expectancy and determination of fixed-term or lifetime annuity price \citep[see, e.g.,][and references therein]{SH20}. The purchase of an annuity is not only a way of guaranteeing one's income for a fixed or life-long period, but it is also essential to the financial viability of the pension and insurance industries. 

Many statistical methods have been proposed for forecasting age-specific mortality rates \citep[see, e.g.,][for a comprehensive review]{Booth06, BT08, SBH11, BCB23}. Among these methods, \cite{LC92} is considered a benchmark in academia and the insurance industry. It is a projection-pursuit-based approach, where age-specific mortality rates observed over the years are decomposed into the first estimated principal component and its associated scores. This decomposition step only retains the principal component that explains the most variance of the observed data and discards the remaining components with relatively smaller variance. The estimated principal component captures the dependence of data in the (deterministic) age dimension, where its scores model the correlation in the temporal dimension. The \citeauthor{LC92}'s \citeyearpar{LC92} method effectively compresses mortality data to achieve dimension reduction but loses valuable information on detailed mortality patterns as a trade-off. 

The \citeauthor{LC92}'s \citeyearpar{LC92} method has been applied to multiple countries, and it has undergone modifications and generalizations to enhance the accuracy of both point and interval forecasts \citep[see, e.g.,][for a review]{SBH11, BCB23}. A common feature of the \citeauthor{LC92}'s \citeyearpar{LC92} method and its variants is that a static principal component analysis is used. By static, principal component analysis is designed to analyze a data matrix observed at a particular time. Because the age-specific mortality rates are observed over the years, they possess strong temporal dependence that should be incorporated into the dimension reduction. 

By treating age as a continuous variable, we consider a functional data-analytic approach of \cite{RS06}. At a high level, characteristics of functional data include being 
\begin{enumerate*}[label=(\roman*)]
    \item high-dimensional,
    \item temporal or structural in nature, and 
    \item recorded over a continuous domain.
\end{enumerate*}
In contrast to traditional statistics, where data are typically represented as a set of discrete observations, functional data analysis considers the function as a whole as the primary unit of study. Considering the data in this way enables the extraction of valuable information regarding the overall shape, trends, and patterns exhibited in the data \citep{JS06}.

\cite{Shang19} extracts dynamic functional principal components by applying eigen-decomposition to an estimated long-run covariance. The estimated long-run covariance includes sample variance and auto-covariance at positive and negative lags. The estimate of long-run covariance can be expressed as the sum of sample autocovariance functions and is often truncated at some finite lag in practice. Using 24 mainly developed countries in the \cite{HMD23}, \cite{Shang19} demonstrated that dynamic functional principal components are superior basis functions for summarizing data features and can achieve improved forecast accuracy compared to their static counterparts based on variance alone. In Section~\ref{sec:3}, we revisit the dynamic functional principal component analysis to extract dynamic principal components and their associated scores. The age-specific mortality forecasts can be obtained through a regression model, and their forecast accuracy can be compared with some holdout samples. 

In addition to the presence of temporal dependence in the time series of age-specific mortality, we also face the challenge of nonstationarity in our data. Because age-specific mortality rates decrease over the years, the mean of the data changes over time, reflecting nonstationary. In this case, \cite{MGG22} suggested taking first-order differencing of functional time series to achieve stationarity. In Section~\ref{sec:4}, we follow \cite{MGG22} and implement the dynamic principal component analysis with the differenced data series to extract dynamic functional principal components and their associated scores, as well as fitted values. After obtaining the functional residuals by subtracting the fitted values from the original functional time series, we conduct an independent test of \cite{GK07} for the residual curves. If the residuals are not independent, we estimate the long-run covariance of the residual process. This allows us to derive additional dynamic principal components and their corresponding scores. This proposed method can be viewed as a two-stage dynamic principal component analysis.

A parallel line of research adopts the functional equivalent of the Beveridge-Nelson decomposition \citep{BN1981} to identify the nonstationary and stationary components (also known as permanent and transitory components) of the nonstationary functional time series \citep[see, e.g.,][]{CHP23, SS24}. A challenge, however, is to determine the nonstationary and stationary subspaces adequately. In an early work by \cite{GS17}, the authors used the vector error correction models to fit a pair of principal component scores of two populations. In the VECM, each set of scores is nonstationary, but a linear combination between the two sets of scores may exist that is stationary in the long run. \cite{LRS23} proposed a general framework for determining the dimensionality of the asymptotically dominant subspace of fractionally integrated functional time series. For a $I(0)$ or $I(1)$ process, the two-stage procedure proposed in this paper is sufficient to address the nonstationarity. For higher-order process $I(2)$, we point out a caveat that differencing alone may not be sufficient. Frequency domain-based inference methods \citep[see, e.g.,][]{VE18} can be the alternative options for such nonstationary functional time series. In contrast to the common practice of differencing $I(1)$ univariate time series before producing forecasts, \cite{choi2020differencing} obtained improved forecasts based on nondifferenced data. However, the idea of using nondifferenced nonstationary data has not yet been explored in functional data analysis. The effectiveness of differencing mortality functional time series data is illustrated in the online supplementary file.

When forecasting mortality, we may come across an additional difficulty: determining the appropriate timeframe of historical data to consider as pertinent for the forecasts. When dealing with a relatively long time series, the data from the distant past have a limited influence on the forecasts. In Section~\ref{sec:5}, we apply weights in the dynamic principal component decomposition to allow more recent data to affect the forecasts more than the data from the distant past \citep[see also][]{HS09}. This is particularly important in mortality, where we can have over 200 years of data, such as Swedish age-specific mortality data in Section~\ref{sec:2}. The data from the 18\textsuperscript{th} and 19\textsuperscript{th} centuries may have poor data quality and are not useful in determining the most accurate principal components for forecasting. By utilizing geometrically decaying weights, we allocate minimal or negligible importance to data from the distant past while assigning relatively greater significance to the most recent data. The weights decrease back in time, and the decaying speed is governed by one tuning parameter, which can be empirically tuned based on a minimum forecast error within a validation set. Section~\ref{sec:7} compares the point and interval forecast accuracy among the standard and weighted two-stage dynamic principal component regression models. 

While the point forecast is considered the best estimate, the interval forecast reflects forecast uncertainty. We introduce a bootstrap procedure to construct prediction intervals by extending the recent work of \cite{PS23} to the nonstationary functional time series. The procedure uses a general vector autoregressive representation of the time-reversed series of principal component scores. It generates backward-in-time, functional replicates that adequately mimic the dependence structure of the underlying process in a model-free way and have the same conditionally fixed curves at the end of each functional pseudo-time series. For this procedure to work, it requires stationarity of the functional time series. Using the idea of \cite{MGG22}, we apply a nonparametric bootstrap procedure in Section~\ref{sec:6} to quantify forecast uncertainty associated with a nonstationary functional time series. Section~\ref{sec:8} concludes with thoughts on how the methods developed here might be further extended.

\section{Swedish age-specific mortality rates}\label{sec:2}
\renewcommand{\theequation}{2.\arabic{equation}}
\setcounter{equation}{0}

We consider Swedish age- and sex-specific mortality rates from 1751 to 2022, obtained from the \cite{HMD23}. Sweden has the longest annual records of age-specific mortality rates among the 41 countries covered by the \cite{HMD23}. We chose this particular time series to demonstrate that more recent data can affect the forecasts more than data from the distant past, hence the necessity of including geometrically decaying weights. The mortality rates are the ratio of death counts per calendar year to population exposure (i.e., mid-year populations) in the relevant intervals of age and time \citep[see][for details]{WAJ+20}. Note that the oldest data for Sweden have lower quality, and \cite{HMD23} advised to use caution when considering data from 1751 to 1860. The cautionary note does not limit the usefulness of the proposed method by putting forward a data-driven way of selecting the adequate length of training samples.

\begin{figure}[!htb]
\centering
\subfloat[Univariate time series plot]
{\includegraphics[width=8.3cm]{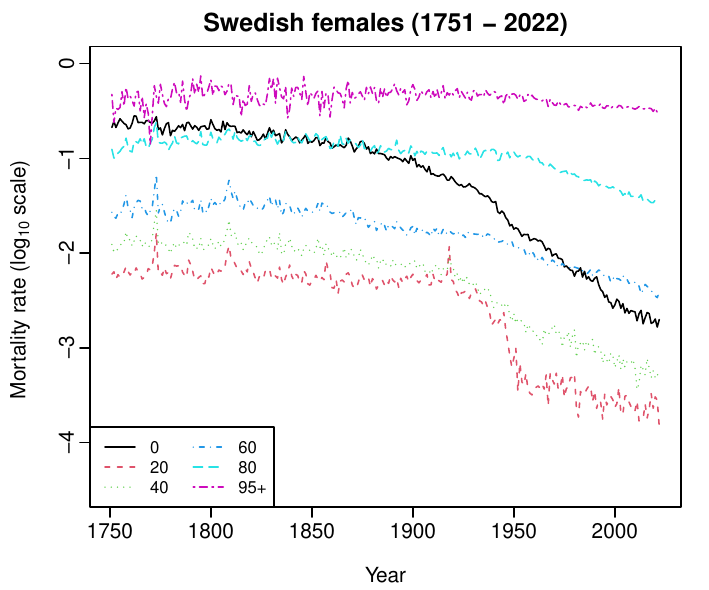}\label{fig:1a}}
\quad
\subfloat[Univariate time series plot]
{\includegraphics[width=8.3cm]{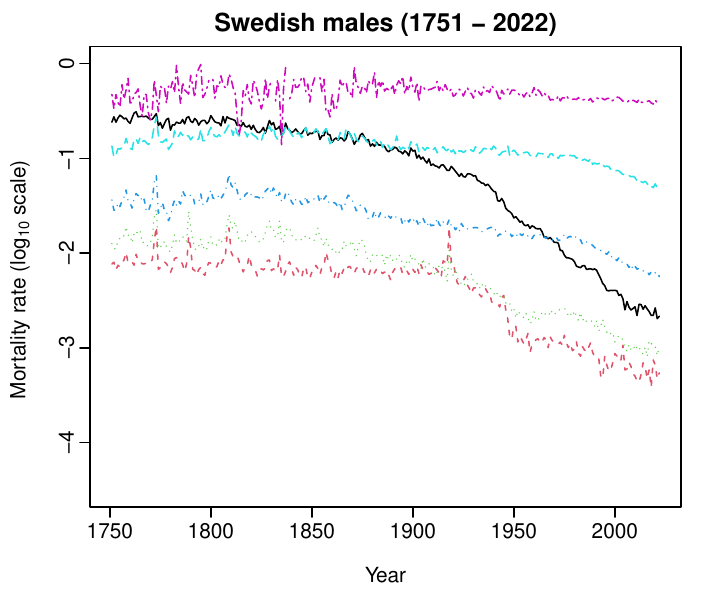}\label{fig:1b}}
\\
\subfloat[Functional time series plot]
{\includegraphics[width=8.3cm]{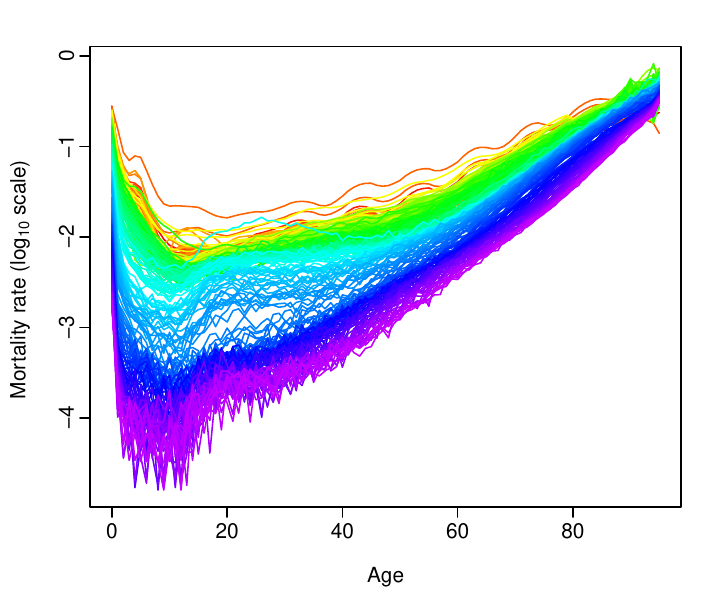}\label{fig:1c}}
\quad
\subfloat[Functional time series plot]
{\includegraphics[width=8.3cm]{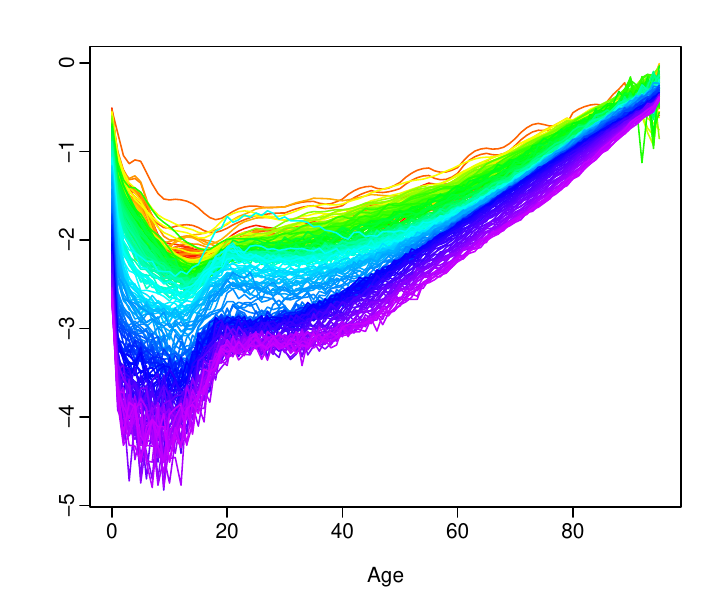}\label{fig:1d}}
\caption{\small Univariate time series plots for Swedish female mortality rates from 1751 to 2022 for ages 0, 20, 40, 60, 80, and 95+ (first row). We adopt a rainbow color palette, displaying the most recent curves in violet, while the curves from the distant past are shown in red (second row).}\label{fig:1}
\end{figure}

Figures~\ref{fig:1a} and~\ref{fig:1b} display the female and male mortality rates in $\log_{10}$ scale for ages 0, 20, 40, 60, 80, and 95+, viewed as a univariate time series. To avoid problems associated with erratic rates at very old ages, we follow the standard practice of grouping the data at ages 95 and older into a single age ``95+" \citep[see, e.g.,][]{HZS21}. Rainbow plots of \cite{HS10} are presented in Figures~\ref{fig:1c} and~\ref{fig:1d}, where the colors of the curves follow the order of a rainbow, with the oldest data shown in red and the most recent data shown in violet. The figures suggest that the Swedish mortality rates gradually declined over the years. In particular, the first row of Figure~\ref{fig:1} indicates a steady and fast drop in infant mortality since the start of the 20th century \citep{Kohler91} and a quick decline in old age mortality \citep{Ledberg20}. The second row of Figure~\ref{fig:1} shows that mortality rates dip from their early childhood high, climb in the teen years, stabilize in the early 20s, and steadily increase with age after 45. 

\section{Modeling stationary functional time series}\label{sec:3}
\renewcommand{\theequation}{3.\arabic{equation}}
\setcounter{equation}{0}

It is commonly assumed that random functions are sampled from a second-order stochastic process $\X$ in $\mathcal{L}^2$, where $\mathcal{L}^2$ is the space of square-integrable functions on the compact set $\mathcal{I}$. Each realization $\X_t$ satisfies the condition $\|\X_t\|^2 = \int_{\mathcal{I}} \X_t^2(u)du<\infty$ with a function support range~$\mathcal{I}$. This space is a separable Hilbert space, with inner product $\langle \X, \Y \rangle=\int_{\mathcal{I}}\X(u)\Y(u)du$ for any two functions, $\X$ and $\Y\in\mathcal{L}^2(\mathcal{I})$ and induced squared norm $\|\cdot\| = \langle \cdot, \cdot \rangle$. The random process $\X \in \mathcal{L}^2$ has well-defined mean and variance \citep[see, e.g., pages 21-23 of][for more details]{horvath2012inference}. In particular, all considered random functions are defined on a common probability space $(\Omega, A, P)$ with sigma operator $A$ and measure $P$. The notation $\X\in \mathcal{L}^p(\Omega, A, P)$ is used to indicate that for some $p>0$, the condition $\text{E}(\|\X\|^p)<\infty$. When $p = 1$, the random process $\X$ has $\text{E}\| \X\| < \infty$. For every unique $\X$, the $\text{E}\langle \cdot, \X \rangle$ is a bounded linear functional in the Hilbert space. Therefore, by the Riesz representation theorem $\text{E}(\X)$ exists and is denoted by $\mu =\text{E}(\X)$. When $p=2$, $\X$ has a linear, self-adjoint, positive semidefinite covariance operator $C(u,v) = \text{Cov}[\X(u), \X(v)] = \text{E}\{[\X(u) - \mu(u)][\X(v) - \mu(v)]\}$, where $u, v\in\mathcal{I}$ \citep[][p.166]{Weidmann80}. 

\subsection{Static functional principal component analysis}\label{sec:3.1}

With the covariance function, via Mercer's lemma, there exists an orthonormal sequence $(\phi_k)$ of continuous function in $\mathcal{L}^2(\mathcal{I})$ and a non-increasing sequence of positive numbers $\theta_k$, such that
\[
C(u,v) = \sum^{\infty}_{k=1}\theta_k \phi_k(u) \phi_k(v),
\]
where the orthonormal functions $[\phi_1(u), \phi_2(u),\dots]$ are referred as the functional principal components. Via the inner product, we can project a time series of functions $\X_t(u)$ onto a collection of orthogonal functional principal components $\phi_k$. This leads to the Karhunen-Lo\`{e}ve expansion
\begin{align*}
\X_t(u) &= \mu(u) + \sum^{\infty}_{k=1}\beta_{t,k}\phi_k(u) \\
&\approx \mu(u) + \sum^K_{k=1}\beta_{t,k}\phi_k(u)
\end{align*}
where $\mu(u)$ denotes the functional mean and $K<n$ denote the number of retained components. The principal component score, $\beta_{t,k}$, is given by the projection of $\X_t^c(u) = \X_t(u) - \mu(u)$ in the direction of the $k$\textsuperscript{th} eigenfunction $\phi_k(u)$, i.e., $\beta_{t,k} = \langle  \X_t^c(u), \phi_k(u)\rangle = \int \X_t^c(u) \phi_k(u) du$.

With a time series of functions $[\X_1(u),\dots,\X_n(u)]$, the functional mean is estimated empirically by $\overline{\X}(u) = \frac{1}{n}\sum^n_{t=1}\X_t(u)$. Via eigen-decomposition, we obtain a set of eigenfunctions $[\widehat{\phi}_1(u),\dots,\widehat{\phi}_K(u)]$, from which the $k$\textsuperscript{th} set of the principal component scores $\widehat{\beta}_{t,k}=\langle \X_t^c(u), \widehat{\phi}_k(u) \rangle$ can be estimated.

\subsection{Estimation of long-run covariance function}\label{sec:3.2}

Let $\{\X_t(u)\}_{t\in \mathbb Z}$ be a stationary and ergodic functional time series. For example, $\X_t(u)$ could be a stock return on the day $t$ at intraday time $u$ or the mortality rate in year $t$ at age $u$. If $\X_t(u)$ is nonstationary, it could be suitably transformed, such as taking the first-order difference, so the stationarity assumption holds. For a stationary functional time series, the long-run covariance is defined as
\begin{align*}
C(u, v) &= \sum^{\infty}_{\ell=-\infty}\gamma_{\ell}(u, v) \\
&=\sum^{\infty}_{\ell=-\infty}\text{cov}[\X_0(u), \X_{\ell}(v)].
\end{align*}
Because $\gamma_{\ell}(u,v)$ is symmetric and nonnegative definite for any $\ell \in \mathbb{Z}$, $C(u, v)$ is also symmetric and nonnegative definite. We obtain a set of eigenvalues and their associated eigenfunctions by applying eigendecomposition to the long-run covariance function, $C(u, v)$.

We resort to a smooth periodogram estimator of \cite{RS17}, which uses a kernel weighting scheme to estimate the long-run covariance. The estimator can be defined as
\begin{equation}
\widehat{C}_{\eta,q}(u, v) = \sum^{\infty}_{\ell=-\infty}W_q\left(\frac{\ell}{\eta}\right)\widehat{\gamma}_{\ell}(u, v), \label{eq:RS_estimator}
\end{equation}
where $\eta$ is a bandwidth parameter, and 
\begin{align*}
   \widehat{\gamma}_\ell(u, v)=\left\{
     \begin{array}{lr}
      \displaystyle \frac{1}{n}\sum_{t=1}^{n-\ell}\left[\X_t(u)-\overline{\X}(u)\right]\left[\X_{t+\ell}(v)-\overline{\X}(v)\right],\quad &\ell \ge 0
      \vspace{.3cm} \\
     \displaystyle \frac{1}{n}\sum_{t=1-\ell}^{n}\left[\X_t(u)-\overline{\X}(u)\right]\left[\X_{t+\ell}(v)-\overline{\X}(v)\right],\quad &\ell < 0,
     \end{array}
   \right.
\end{align*}
is an estimator of $\gamma_{\ell}(u, v)$, and $W_q(\cdot)$ is a symmetric weight function with bounded support of order $q$. The performance of the smooth periodogram estimator depends on the bandwidth parameter $\eta$, and we use the plug-in bandwidth selection algorithm of \cite{RS17} to select the optimal bandwidth parameter. This plug-in bandwidth selection algorithm has also been used in \cite{MGG22}.

\subsection{Dynamic functional principal component analysis}

With the estimated long-run covariance in~\eqref{eq:RS_estimator}, we apply a principal component decomposition to extract dynamic principal components and their associated scores. Via Mercer's lemma, the estimator of the long-run covariance function can be expressed as
\[
\widehat{C}_{\eta,q}(u,v)=\sum^{\infty}_{k=1}\widehat{\lambda}_k \widehat{\phi}_k(u)\widehat{\phi}_k(v)
\]
where $\widehat{\lambda}_1>\widehat{\lambda}_2>\ldots>0$ are the estimated eigenvalues of $\widehat{C}_{\eta,q}(u,v)$, and $[\widehat{\phi}_1(u),\widehat{\phi}_2(u),\ldots]$ are the estimated functional principal components. The principal component score, $\widehat{\beta}_{t,k}$, is given by the projection of $\X_t^c(u)$ in the direction of the $k$\textsuperscript{th} eigenfunction $\widehat{\phi}_k(u)$.

A realization of the stochastic process, $\X_t(u)$, can be written as
\begin{equation}
\X_t(u) = \overline{\X}(u) + \sum^K_{k=1}\widehat{\beta}_{t,k}\widehat{\phi}_k(u)+e_t(u),\qquad t=1,2,\dots,n, \label{eq:3.2}
\end{equation}
where $e_t(u)$ denotes a residual function. Several hypothesis tests, including the independent test of \cite{GK07}, have been developed based on $[e_1(u),\dots,e_n(u)]$.

Following \cite{AH13}, \cite{LRS20} and \cite{MGG22}, the value of $K$ is determined as the integer minimizing ratio of two adjacent empirical eigenvalues given by
 \[
\widehat{K} = \argmin_{1\leq k\leq k_{\max}}\left\{\frac{\widehat{\lambda}_{k+1}}{\widehat{\lambda}_k}\times \mathds{1}\left(\frac{\widehat{\lambda}_k}{\widehat{\lambda}_1}\geq \delta\right)+\mathds{1}\left(\frac{\widehat{\lambda}_k}{\widehat{\lambda}_1}<\delta\right)\right\},
\]
where $k_{\max}$ is a pre-specified positive integer, $\delta$ is a pre-specified small positive number, and $\mathds{1}(\cdot)$ is the indicator function. We choose $k_{\max} = \#\{k|\widehat{\lambda}_k\geq \sum^n_{k=1}\widehat{\lambda}_k/n, k\geq 1\}$ and set the threshold constant $\delta = 1/\ln(\max\{\widehat{\lambda}_1,n\})$.  

There are other approaches for selecting $K$ that are commonly used in empirical applications: 
\begin{enumerate*}[label=(\arabic*)]
  \item the proportion of total variance explained by the first few leading functional principal components \citep{Chiou2012,shang2017grouped};
  \item Akaike information criterion \citep{yao2005functional};
  \item bootstrap methods \citep{hall2006assessing};
  \item predictive cross-validation leaving out one or more curves \citep{liu2022predictive}.
\end{enumerate*}
Although choosing the number of components in the implementation of functional principal component analysis (FPCA) can be a research topic on its own, it is beyond the scope of this article.

\subsection{Dynamic principal component regression}\label{sec:3.4}

Conditioning on the estimated mean function $\overline{\X}(u)$, the set of estimated dynamic principal components $\bm{\widehat{\Phi}}(u)=\{\widehat{\phi}_1(u), \widehat{\phi}_2(u), \dots,\widehat{\phi}_{\widehat{K}}(u)\}$ and observed data $\bm{\X}(u)$, we can obtain $h$-step-ahead forecasts of $\X_{n+h}(u)$:
\begin{align}
\X_{n+h|n}(u) &= \E[\X_{n+h}(u)|\overline{\X}(u),\bm{\widehat{\Phi}}(u), \bm{\X}(u)] \notag\\
&=\overline{\X}(u)+\sum^{\widehat{K}}_{k=1}\widehat{\beta}_{n+h|n,k}\widehat{\phi}_k(u),
\end{align}
where $\widehat{\beta}_{n+h|n,k}$ denotes the $h$-step-ahead forecast of $\beta_{n+h,k}$ obtained via a univariate time series forecasting method. 

Among the univariate time series forecasting methods, the common ones are the exponential smoothing (ETS) and autoregressive integrated moving average (ARIMA) models. Since the yearly age-specific mortality rates do not contain seasonality, the ARIMA has the general form of
\[
(1-\tau_1 B - \cdots - \tau_p B^p)(1-B)^d\bm{\beta}_k=\psi + (1+\nu_1B + \cdots + v_q B^q)\bm{w}_k,
\]
where $\psi$ represents the intercept, $(\tau_1,\dots,\tau_p)$ denote the coefficients associated with the autoregressive component, $\bm{\beta}_k = \{\beta_{1,k},\dots,\beta_{n,k}\}$ represents principal component scores, $(\nu_1,\dots, \nu_q)$ denote the coefficients associate with the moving average component, $B$ denotes the backshift operator, $d$ denotes the differencing operator, and $\bm{w}_k = \{w_{1,k},\dots,w_{n,k}\}$ represents a white-noise error term. 

An automatic algorithm of \cite{HK08} selects the optimal autoregressive order~$p$, moving average order~$q$, and difference order~$d$. The value of $d$ is selected based on successive Kwiatkowski-Phillips-Schmidt-Shin (KPSS) unit root tests \citep{KPSS92}. KPSS tests are used to test the null hypothesis that an observable time series is stationary around a deterministic trend. We first test the original time series for a unit root; if the test result is significant, we test the differenced time series for a unit root. The procedure continues until we obtain our first insignificant result. Once the order d is determined, the orders of $p$ and $q$ are selected based on the optimal AIC with a correction for a small finite sample size \citep{Akaike74}. Having identified the optimal ARIMA model, the maximum likelihood method can be used to estimate the parameters.

For comparison, we also consider a multivariate time series forecasting method, namely the vector autoregressive of order $p$ model. The order $p$ is selected based on the corrected Akaike information criterion introduced in \cite{HT93}. In Section~\ref{sec:7.5}, we evaluate and compare the forecast accuracy between the univariate ARIMA and multivariate VAR models.

\section{Modeling nonstationary functional time series}\label{sec:4}
\renewcommand{\theequation}{4.\arabic{equation}}
\setcounter{equation}{0}

In the functional time series literature, many methods were designed for modelling and forecasting stationary series, and its extension to nonstationary series is challenging yet important. Without stationarity, the long-run covariance function is not well-defined. A solution is to take first-order differencing of the original functional time series to achieve stationarity. The differenced series allows us to estimate the long-run covariance function and extract dynamic functional principal components. Diagnostic checks, such as the independent test of \cite{GK07}, can be performed from the residual functions. If the residuals are still temporally correlated, we perform the second-stage functional principal component decomposition to extract the temporal patterns remaining in the residuals. A set of estimated principal component scores is obtained in each dynamic functional principal component decomposition. Conditional on the observed data, the estimated mean function, and estimated functional principal components, the $h$-step-ahead forecast functional time series can be obtained from the forecast principal component scores. We demonstrate the superiority of the two-stage method over the traditional FPCA in producing more accurate point forecasts in the online supplementary file.

Following early work by \cite{MGG22}, we extend the functional principal component regression from stationary to nonstationary functional time series. By nonstationarity, it is manifested through at least one of the principal component scores $(\widehat{\beta}_{t,1},\widehat{\beta}_{t,2},\dots,\widehat{\beta}_{t,\widehat{K}})$ where $\widehat{K}$ is obtained from the eigenvalue ratio criterion. We assume that the nonstationary principal component scores exhibit an $I(1)$ scalar-valued process. Let $r\in \{1,2,\dots,\widehat{K}\}$ be the number of principal component scores that are $I(1)$ processes. Although it is possible to have $r=\widehat{K}$, it is more common to observe $r<\widehat{K}$ and the remaining $(\widehat{K} - r)$ process are stationary \citep[see also][]{CKP16}. In the latter case, a part of the underlying process resides in the nonstationary Hilbert space, while the remaining part resides in the stationary Hilbert space \citep[see also][]{SS24}. 

To estimate the dynamic functional principal components, we consider a two-step procedure:
\begin{enumerate}
\item[1)] Compute the estimated long-run covariance based on the first-order differencing of the original functional time series, denoted by $\Delta \X_t(u) = \X_t(u)-\X_{t-1}(u)$ for $t=2,\dots,n$ and let $\widehat{r}$ be the estimated number of principal components, obtained by the eigenvalue ratio criterion.
\item[2)] With the estimated \textcolor{red}{$\widehat{r}$}, we compute the estimated functional principal components and their associated scores, $\{\widehat{\zeta}_k\}$ and $\{\widehat{\beta}_{t,k}=\langle \X_t(u), \widehat{\zeta}_k(u)\rangle\}$ for $k=1,2,\dots, \widehat{r}$ and $t=1,2,\dots,n$.
\item[3)] Compute the functional residuals: $\Z_t(u) = \X_t(u) - \overline{\X}(u)  - \sum_{k=1}^{\textcolor{red}{\widehat{r}}}\widehat{\beta}_{t,k}\widehat{\zeta}_{k}(u)$.
\item[4)] Apply the independence test of \cite{GK07} to $\bm{\Z}(u) = [\Z_1(u),\dots,\Z_n(u)]$. If $\bm{\Z}(u)$ is independent, terminate; Else, compute the estimated long-run covariance function of $\bm{\Z}(u)$.
\item[5)] Apply the eigenvalue ratio criterion to select the retained number of principal components $\widehat{K}-\textcolor{red}{\widehat{r}}$.
\item[6)] Obtain all estimated functional principal components $\{\widehat{\zeta}_{\omega}(u), \omega = \widehat{r}+1,\dots,\widehat{K}\}$ and their principal component scores $\widehat{\beta}_{t,\omega} = \langle \X_t(u), \widehat{\zeta}_{\omega}(u)\rangle$ for $\omega = 1,2,\dots,(\widehat{K}-\textcolor{red}{\widehat{r}})$.
\end{enumerate}

\section{Geometrically decaying weights}\label{sec:5}
\renewcommand{\theequation}{5.\arabic{equation}}
\setcounter{equation}{0}

After taking the first-order differencing of the original functional time series, we apply a set of geometrically decaying weights to obtain weighted data. It can be expressed as
\[
\Delta^{*} \X_t(u)  = \sum^n_{t=1}w_t\Delta \X_t(u), 
\]
where $w_t = \kappa(1-\kappa)^{n-t}$ is a geometrically decreasing weight with $0<\kappa<1$, and $\sum_{t=1}^n w_t = 1$. We defer the selection of $\kappa$ to Section~\ref{sec:7.1}.

Computationally, we discretize $\X_t(u)$ on a dense grid of $p$ equally spaced points $(u_1, u_2, \dots, u_p)$ that spans the function support range. Denote the discretized $\Delta \X_t(u)$ as an $(n-1)\times p$ matrix $\bm{G}^*$ and let $\bm{G}=\bm{W}\bm{G}^*$, where $\bm{W}=\text{diagonal}(w_1, w_2,\dots,w_{n-1})$. Applying singular value decomposition to $\bm{G}$ gives $\bm{G} = \bm{\Psi}\bm{\Lambda}\bm{V}^{\top}$, where $\widehat{\zeta}_k^*(u_j)$ is the $(k, j)$\textsuperscript{th} element of $\bm{\Psi}$. Then, the $k$\textsuperscript{th} set of the estimated principal component scores is obtained by
\begin{equation*}
\widehat{\beta}^*_{t,k} = \langle \X_t(u) - \overline{\X}(u), \widehat{\zeta}^*_k(u)\rangle.
\end{equation*}
With the estimated functional principal components and their scores, we compute the functional residuals
\begin{equation*}
\Z_t^*(u) = \X_t(u) - \overline{\X}(u) - \sum^{\widehat{r}}_{k=1}\widehat{\beta}^{*}_{t,k}\widehat{\zeta}^{*}_{k}(u).
\end{equation*}
Since $\Z_t^*(u)$ models the functional residual process, assigning another set of geometrically decaying weights is unnecessary. The remaining procedure is the same as the standard approach.

\section{Construction of prediction intervals via bootstrapping} \label{sec:6}
\renewcommand{\theequation}{6.\arabic{equation}}
\setcounter{equation}{0}

Prediction intervals are valuable for assessing the probabilistic uncertainty associated with point forecasts. The forecast uncertainty stems from systematic deviations (e.g., due to parameter or model uncertainty) and random fluctuations (e.g., due to model error term). As was emphasized by \cite{Chatfield93, Chatfield00}, it is essential to provide interval forecasts as well as point forecasts to
\begin{enumerate}
\item[(1)] assess future uncertainty levels;
\item[(2)] enables different strategies to be planned for a range of possible outcomes indicated by the interval forecasts;
\item[(3)] compare forecasts from different methods more thoroughly; and 
\item[(4)] explore different scenarios based on various assumptions.
\end{enumerate}

We consider two sources of uncertainty: truncation errors in the functional principal component decomposition and forecast errors in the forecast principal component scores. Since principal component scores are regarded as surrogates of the original functional time series, these principal component scores capture the temporal dependence structure inherited in the original functional time series \citep[see also][]{Paparoditis18, Shang18}. By adequately bootstrapping the forecast principal component scores, we can generate a set of bootstrap $\bm{\X}_{n+h}^*$, conditional on the estimated mean function and estimated functional principal components from the observed $(\X_1,\X_2,\dots,\X_n)$.

Using a univariate time series model, we can obtain multi-step-ahead forecasts for the principal component scores, $\{\widehat{\beta}_{1,k},\dots,\widehat{\beta}_{n,k}\}$ for $k=1,\dots, \widehat{r}$. Let the $h$-step-ahead forecast errors be given by $\widehat{\varsigma}_{t,h,k}=\widehat{\beta}_{t,k}-\widehat{\beta}_{t|t-h,k}$ for $t=h+1,\dots,n$. These errors can then be sampled with replacement to give bootstrap samples of $\beta_{n+h,k}$:
\begin{equation*}
\widehat{\beta}_{n+h|n,k}^{(b)} = \widehat{\beta}_{n+h|n,k}+\widehat{\varsigma}_{*,h,k}^{(b)},\quad b=1,\dots,B,
\end{equation*}
where $B=1,000$ symbolizes the number of bootstrap replications and $\widehat{\varsigma}_{*,h,k}^{(b)}$ are sampled with replacement from $\{\widehat{\varsigma}_{t,h,k}\}$.

Similarly, we can also obtain multi-step-ahead forecasts for the principal component scores, $\{\widehat{\beta}_{1,\omega},\dots,\widehat{\beta}_{n,\omega}\}$ for $\omega=1,\dots,(\widehat{K}-\widehat{r})$. Denote the $h$-step-ahead forecast errors as $\upsilon_{t,h,\omega}=\widehat{\beta}_{t,\omega} - \widehat{\beta}_{t|t-h,\omega}$. These errors can then be sampled with replacement to give bootstrap samples of $\beta_{n+h,\omega}$:
\begin{equation*}
\widehat{\beta}_{n+h|n,\omega}^{(b)} = \widehat{\beta}_{n+h|n,\omega}+\widehat{\upsilon}_{*,h,\omega}^{(b)}.
\end{equation*}

Assuming the two-step functional principal components approximate the data $\bm{\X}$ relatively well, the model residual 
\begin{equation}
\Y_t(u) = \X_t(u) - \overline{\X}(u) - \sum^{\widehat{r}}_{k=1}\widehat{\beta}_{t,k}\widehat{\zeta}_{k}(u) - \sum^{(\widehat{K}-\widehat{r})}_{\omega=1}\widehat{\beta}_{t,\omega}\widehat{\zeta}_{\omega}(u)  \label{eq:6.1}
\end{equation}
should contribute nothing but random noise. In the weighted dynamic factor model, the estimated basis functions $[\widehat{\zeta}_1(u), \widehat{\zeta}_2(u), \dots,\widehat{\zeta}_{r}(u)]$ are extracted differently. That difference is because the eigen-decomposition was performed on the estimated long-run covariance function. Consequently, we can bootstrap the model fit errors in~\eqref{eq:6.1} by sampling with replacement from the model residual term $\{\Y_1(u),\dots,\Y_n(u)\}$.

Adding all sources of variability, we obtain $B$ variants for $\X_{n+h}(u)$:
\begin{equation*}
\X_{n+h|n}^{(b)}(u) = \overline{\X}(u) + \sum^{\widehat{r}}_{k=1} \widehat{\beta}_{n+h|n,k}^{(b)}\widehat{\zeta}_k(u) + \sum^{(\widehat{K}-\widehat{r})}_{\omega=1} \widehat{\beta}_{n+h|n,\omega}^{(b)}\widehat{\zeta}_{\omega}(u) + \Y^{(b)}(u),
\end{equation*}
where $\widehat{\beta}_{n+h|n,k}^{(b)}$ and $\widehat{\beta}_{n+h|n,\omega}^{(b)}$ denote the bootstrap forecast of the scores obtained from standard and weighted functional principal component analyses. With the bootstrapped $\X_{n+h|n}^{(b)}(u)$, the pointwise prediction intervals are obtained by taking $\alpha/2$ and $(1-\alpha/2)$ quantiles at the $100(1-\alpha)\%$ nominal coverage probability, where $\alpha$ denotes a level of significance, customarily $\alpha=0.2$ or 0.05.

\section{Results}\label{sec:7}

\subsection{Selection of the weight parameter}\label{sec:7.1}

\begin{figure}[!htb]
\begin{center}
\begin{tikzpicture}
\draw (0,0) rectangle (15,2);
\draw (5,2) -- (5,0);
\draw (10,2) -- (10,0);
\draw (2.5,1) node {Training};
\draw (7.5,1) node[red] {Validation}; 
\draw (12.5,1) node {Testing};
\draw(2.5,2.2) node{$1:(n-60)$};
\draw(7.5,2.2) node{$(n-59):(n-30)$};
\draw(12.5,2.2) node{$(n-29):n$};
\end{tikzpicture}
\end{center}
\caption{\small Illustration of the cross-validation method for a sample of size $n$ ($n=272$ for the considered Swedish mortality data set). A model is constructed using data in the training set to forecast data in the validation set. The model's predictive ability is evaluated based on forecast error. The optimal value of $\kappa$ is determined based on the minimal forecast error in the validation set.}\label{fig:2}
\end{figure}
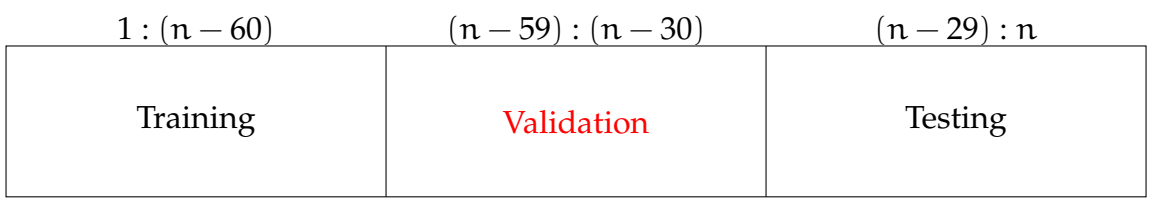

As described in Section~\ref{sec:2}, we consider age- and sex-specific mortality rates in Sweden between 1751 and 2022. We evaluate the forecasts at integer ages between 0 and 95, with the last age corresponding to the grouped data at 95 and older. We apply the weighted penalized regression splines to smooth the raw mortality observations and consider a monotonic constraint for ages over 65 \citep[see][for details]{HU07}. The data samples are split into training, validation, and testing sets in this empirical analysis. Out of the total $n = 272$ annual observations, we use the first 272 years (approximately 80\%) of data as the initial training set, the next 30 years (approximately 10\%) of data as the validation set, and the final 30 years (approximately 10\%) of data as the testing set. To select the weight parameter, we aim to find $\kappa$ that minimizes a forecast error via the \verb| optimize | function in \Rlogo. The forecast error can be the root mean squared prediction error (RMSPE) and mean absolute prediction error (MAPE) described in Section~\ref{sec:7.3}. Alternatively, the interval forecast error can also be the coverage probability difference (CPD) or interval score in Section~\ref{sec:7.4}. In Figure~\ref{fig:2}, we visually display the sample splitting.

\subsection{Selection of functional principal components}\label{sec:7.2}

Using an expanding window approach, we evaluate and compare the forecast accuracy of the standard and weighted dynamic factor models. The expanding window analysis of a time series model is commonly used to assess model and parameter stability over time and prediction accuracy. The expanding window analysis determines the constancy of a model's parameter by computing parameter estimates and their resultant forecasts over an expanding window of a fixed size through the sample \citep[for detail][pp. 313-314]{ZW06}. 

Using the initial 242 observations from 1751 to 1992, we produce one- to 30-step-ahead forecasts. Through the expanding-window approach, we re-estimate the parameters in the time series forecasting models from the first 243 observations from 1751 to 1993. Forecasts from the estimated models are then produced for one- to 29-step-ahead forecasts. We iterate this process by increasing the sample size by one year until the end of the data period in 2022. This process produces 30 one-step-ahead forecasts, 29 two-step-ahead forecasts, $\dots$, and one 30-step-ahead forecast.

\begin{figure}[!htb]
\centering
\subfloat[FPCA in step~1]
{\includegraphics[width=8.7cm]{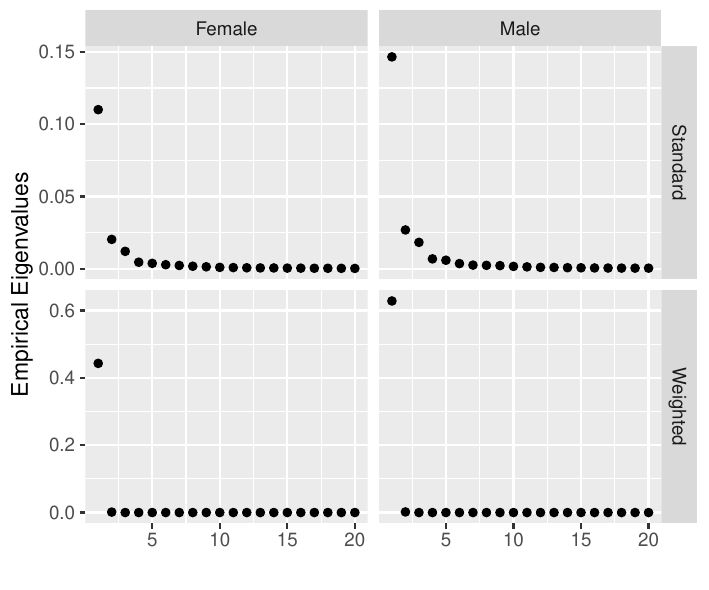}\label{fig:3a}}
\quad
\subfloat[FPCA in step~2]
{\includegraphics[width=8.7cm]{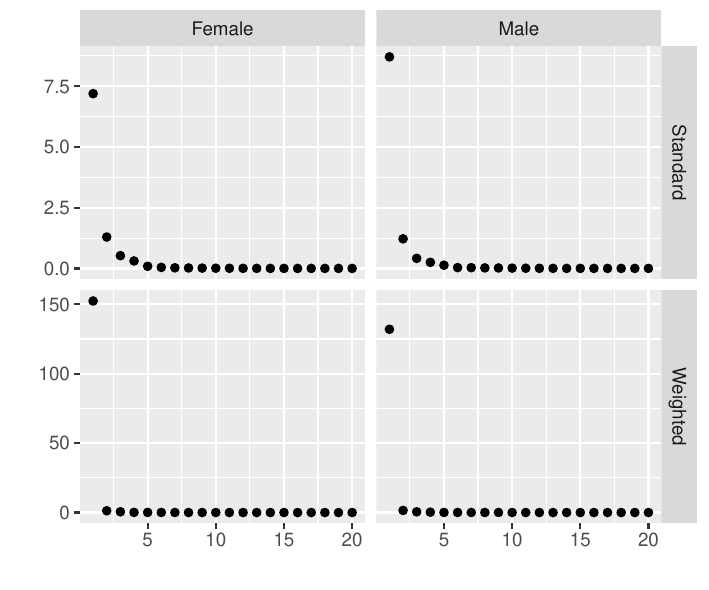}\label{fig:3b}}
\caption{Empirical eigenvalues obtained by implementing the proposed two-step FPCA approach to 271 years of annual Sweden age-specific mortality rates.}\label{fig:3}
\end{figure}

Figure~\ref{fig:3} illustrates the empirical eigenvalues obtained in the last iteration of the expanding window applications. When applied to the annual mortality observations from 1971 to 2021, the proposed two-stage FPCA approach yields $\widehat{r} = 1$, $ \widehat{K} = 2$ and $\widehat{K}-\widehat{r} = 1$ for both genders. The figures indicate that applying the geometrically decaying weights widens the gaps between the first and the second empirical eigenvalues in both steps of FPCA, enhancing the adequate feature extraction from the mortality time series data.

\begin{figure}[!htb]
\centering
\subfloat[Female]
{\includegraphics[width=8.7cm]{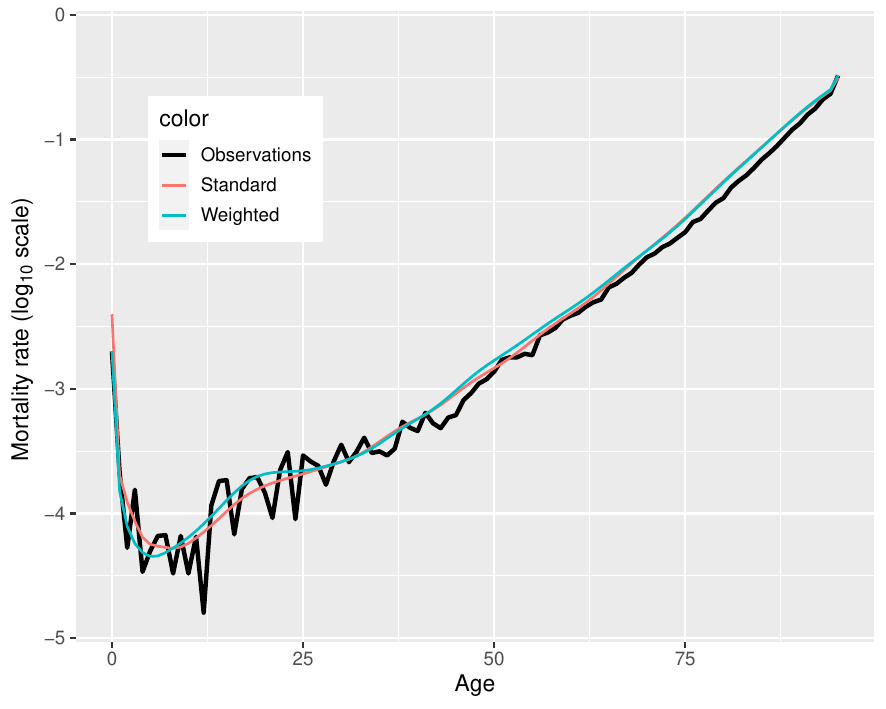}\label{fig:4a}}
\quad
\subfloat[Male]
{\includegraphics[width=8.7cm]{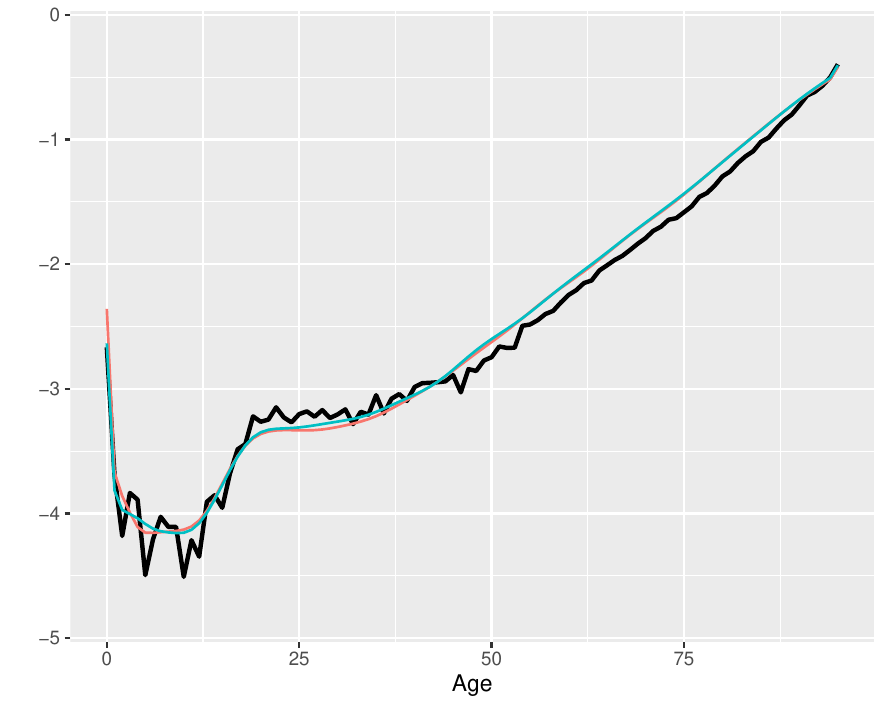}\label{fig:4b}}
\caption{A comparison of age-specific mortality observations and forecasts for Sweden females and males in the year 2022.}\label{fig:4}
\end{figure}

Using the extracted dynamic functional principal components and the ARIMA models, one-step-ahead forecasts for Swedish age-specific mortality rates in 2022 are obtained and shown in Figure~\ref{fig:4}. The weighted dynamic functional factor models appear to give higher forecasts than the standard method for young females between 18 and 25 years old and middle-aged women between 45 and 60 years old. The weighted approach produces slightly higher forecasts than the standard method for males between 25 and 35. More details about forecast comparison between the weighted and standard approach will be discussed in Section~\ref{sec:7.3}.

\subsection{Point forecast results}\label{sec:7.3}

We compare the forecasts with the holdout samples to evaluate the point forecast accuracy and determine the out-of-sample forecast accuracies. Specifically, we consider the RMSPE and MAPE that measures how close the forecasts are compared with the actual values of the variable being forecast, regardless of the direction of forecast errors. The error measure can be written as
\begin{align*}
\text{RMSPE}(h) &= \sqrt{\frac{1}{96\times (31-h)}\sum^{30}_{\xi = h}\sum^{96}_{j=1}[\X_{m+\xi}(u_j) - \widehat{\X}_{m+\xi}(u_j)]^2} \\
\text{MAPE}(h) &= \frac{1}{96\times (31-h)}\sum^{30}_{\xi = h}\sum^{96}_{j=1}\left|\X_{m+\xi}(u_j) - \widehat{\X}_{m+\xi}(u_j)\right|
\end{align*}
where $m=(n-30)$ denotes the last year of the validation sample, and $h=1,2,\dots,30$ denotes a forecast horizon. For each horizon, we consider forecasting each set of the estimated principal component scores by the ARIMA or ETS model \citep[see][for a detailed review]{HKO+08}. 

\begin{figure}[!htb]
\centering
\includegraphics[width=8.5cm]{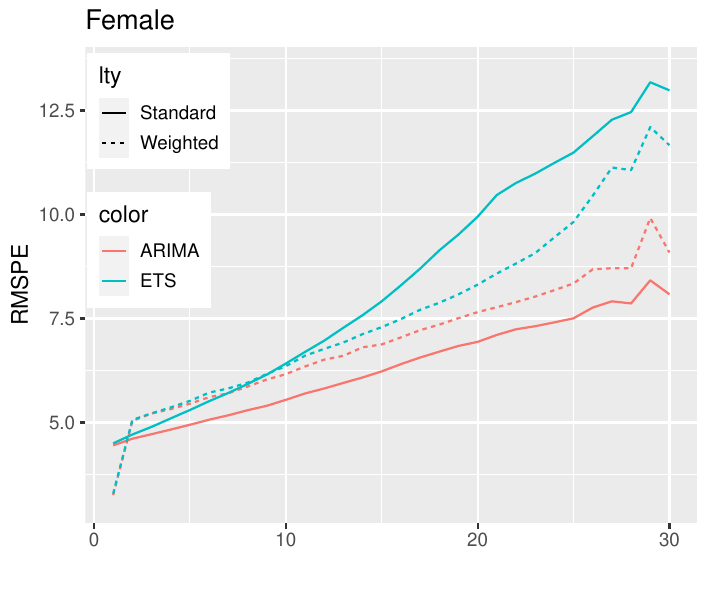}
\quad
\includegraphics[width=8.5cm]{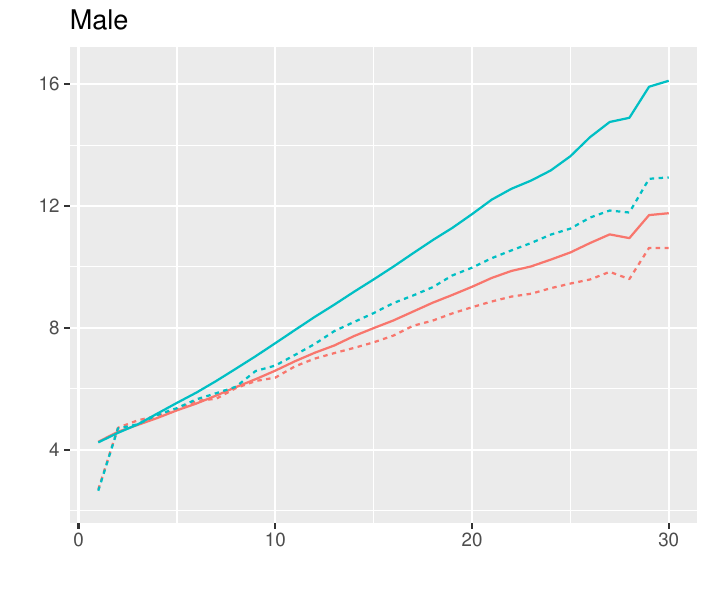}
\\
\includegraphics[width=8.5cm]{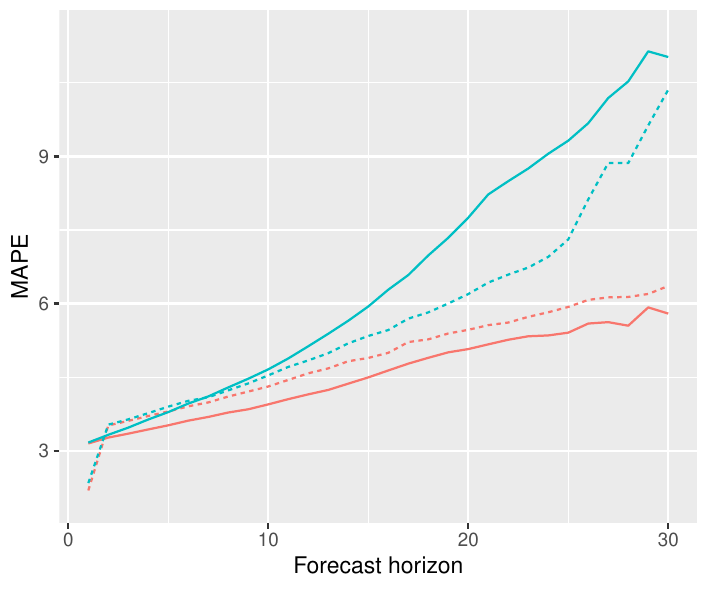}
\quad
\includegraphics[width=8.5cm]{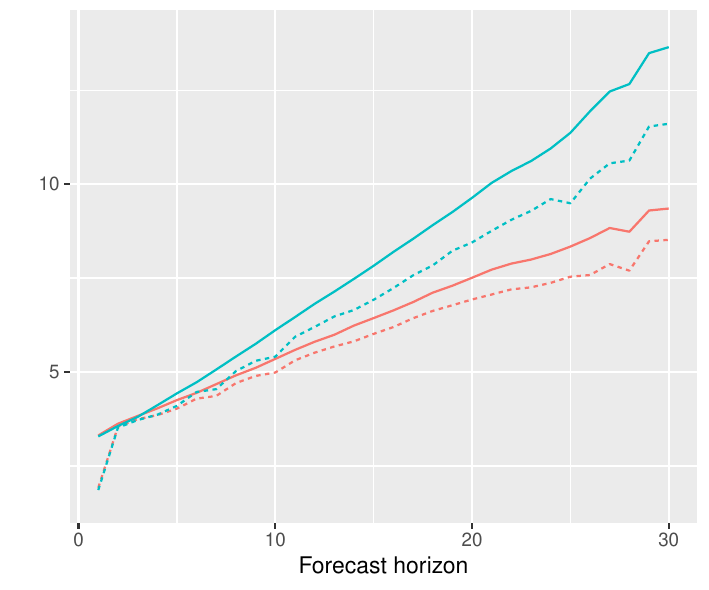}
\caption{\small A comparison of the point forecast accuracy, measured by the RMSPE and MAPE, between the standard and weighted dynamic functional factor models.}\label{fig:5}
\end{figure}

The horizon-specific point forecast error results are presented in Figure~\ref{fig:5}. As the forecast horizon increases, the forecast errors generally become larger. This phenomenon is because the fitted model is no longer optimal as we look into the future. Between the ARIMA and ETS models, the ARIMA model provides smaller forecast errors than the ones from the ETS model. This result may be because the ARIMA model can handle nonstationary series via differencing. Using the ARIMA model, the weighted approach generally produces smaller forecast errors for the male series but not for the female series. Compared with the female series, the male series is difficult to model due to a low signal-to-noise ratio.

\subsection{Interval forecast results}\label{sec:7.4}

To evaluate and compare the interval forecast accuracy, we consider the interval score of \cite{GR07} and \cite{GK14}. For each year in the forecasting period, the $h$-step-ahead prediction intervals are computed at the $100(1-\alpha)\%$ nominal coverage probability. We consider the common case of the symmetric $100(1-\alpha)\%$ prediction intervals, with lower and upper bounds that are predictive quantiles, denoted by $\widehat{\X}_{m+\xi}^{\text{lb}}(u_j)$ and $\widehat{\X}_{m+\xi}^{\text{ub}}(u_j)$. We compute the empirical coverage probability, defined as
\[
\text{Coverage}(h) = 1-\frac{1}{(31-h)\times 96}\sum^{30}_{\xi=h}\sum_{j=1}^{96}\left[\mathds{1}\{\X_{m+\xi}(u_j)>\widehat{\X}_{m+\xi}^{\text{ub}}(u_j)\} + \mathds{1}\{\X_{m+\xi}(u_j)<\widehat{\X}_{m+\xi}^{\text{lb}}(u_j)\}\right],
\]
where $\mathds{1}\{\cdot\}$ represents the binary indicator function, and $(n-29,\dots,n)$ denote the number of curves in the forecasting period. While the empirical coverage probability reveals over- or under-estimation of the nominal coverage probability, it is not an accuracy criterion due to the possible cancellation. As an alternative, the CPD$(h)$ is defined as
\[
\text{CPD}(h) = \left|\text{Coverage}(h) - (1-\alpha)\right|.
\]
The smaller the value of CPD is, the better the method is. Although the empirical coverage probability and CPD are adequate measures, neither do they consider the sharpness of the prediction interval, i.e., the distance between the lower and upper bounds. To rectify this problem, as defined by \cite{GR07} and \cite{GK14}, a scoring rule for the interval forecasts at time point $\X_{m+\xi}(u_j)$ is
\begin{align*}
S_{\alpha,\xi}\left[\widehat{\X}_{m+\xi}^{\text{lb}}(u_j),\widehat{\X}_{m+\xi}^{\text{ub}}(u_j),\X_{m+\xi}(u_j)\right] = &\left[\widehat{\X}_{m+\xi}^{\text{ub}}(u_j),-\widehat{\X}_{m+\xi}^{\text{lb}}(u_j)\right] \\
& + \frac{2}{\alpha}\left[\widehat{\X}_{m+\xi}^{\text{lb}}(u_j)-\X_{m+\xi}(u_j)\right]\mathds{1}\left\{\X_{m+\xi}(u_j)<\widehat{\X}_{m+\xi}^{\text{lb}}(u_j)\right\}\\
& + \frac{2}{\alpha}\left[\X_{m+\xi}(u_j) - \widehat{\X}_{m+\xi}^{\text{ub}}(u_j)\right]\mathds{1}\left\{\X_{m+\xi}(u_j) > \widehat{\X}_{m+\xi}^{\text{ub}}(u_j)\right\}.
\end{align*}
The interval score rewards a narrow prediction interval if and only if the holdout observation lies within the prediction interval. The optimal interval score is achieved when $\X_{m+\xi}(u_j)$ lies between $\widehat{\X}_{m+\xi}^{\text{lb}}(u_j)$ and $\widehat{\X}_{m+\xi}^{\text{ub}}(u_j)$, and the pointwise distance between $\widehat{\X}_{m+\xi}^{\text{lb}}(u_j)$ and $\widehat{\X}_{m+\xi}^{\text{ub}}(u_j)$ is minimal.

\begin{figure}[!htb]
\centering
\includegraphics[width=8.5cm]{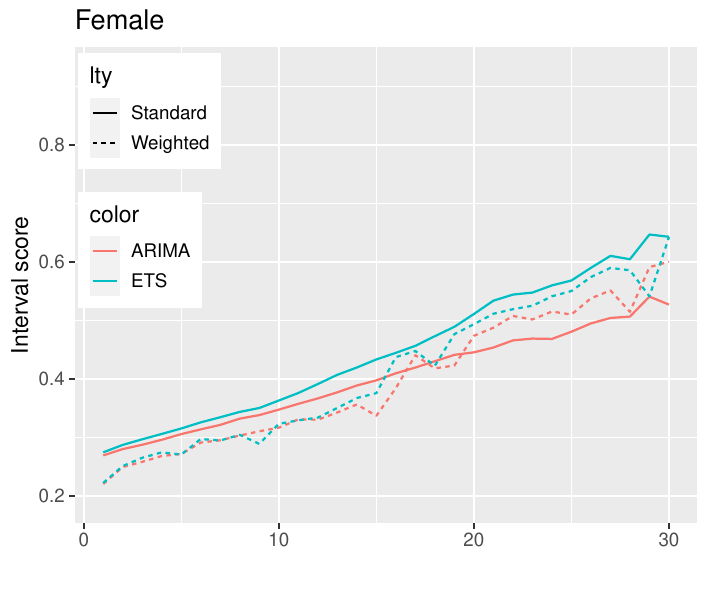}
\quad
\includegraphics[width=8.5cm]{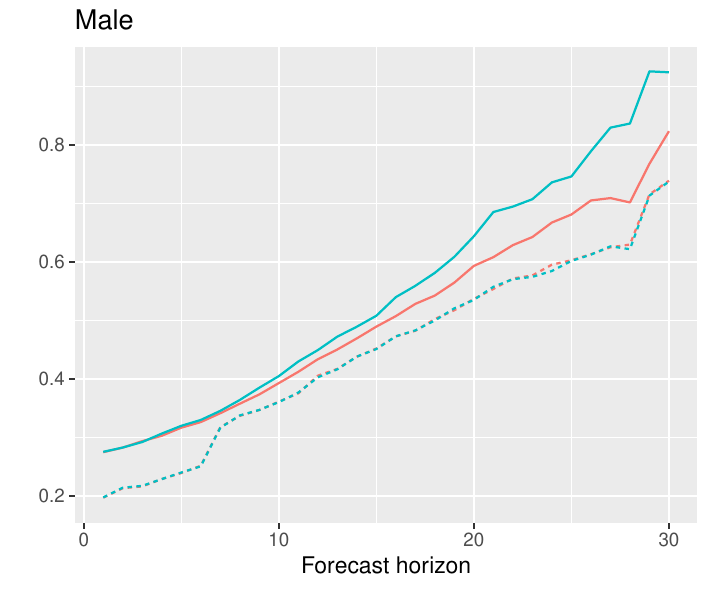}
\\
\includegraphics[width=8.5cm]{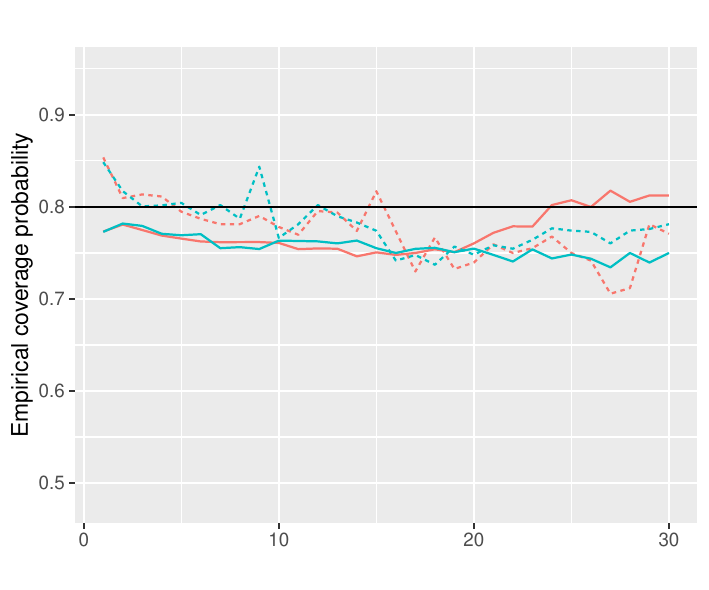}
\quad
\includegraphics[width=8.5cm]{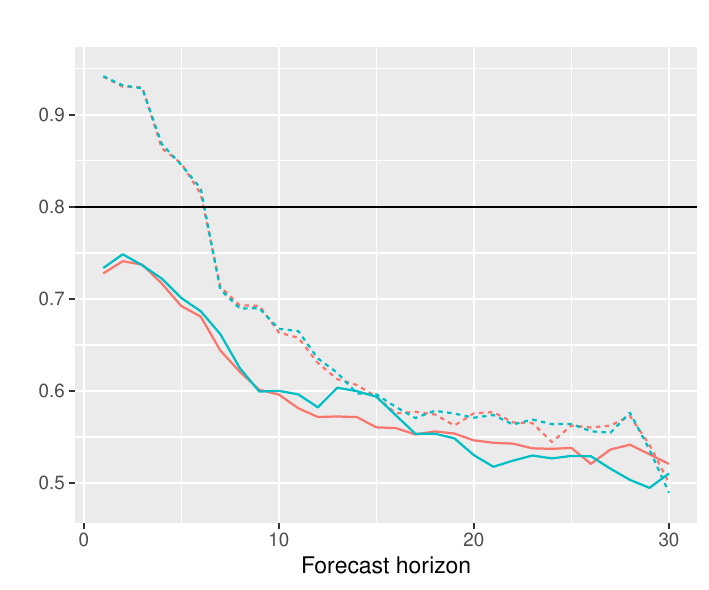}
\caption{\small A comparison of the interval forecast accuracy, measured by the interval score and empirical coverage probability, between the standard and weighted dynamic functional factor models.}\label{fig:6}
\end{figure}

We consider forecasting each set of the estimated principal component scores by the ETS or ARIMA method for each horizon, where $h=1,2,\dots,30$. The horizon-specific interval forecast error results are presented in Figure~\ref{fig:6}. As the forecast horizon increases, the interval scores generally become larger. Further, the empirical coverage probabilities are closer to the nominal one for the female data. For the male data, the empirical coverage probabilities deviate further from the nominal one. Between the ARIMA and ETS models, the ARIMA model provides smaller interval forecast errors than the ones from the ETS model. Using the ARIMA model, the weighted approach generally produces smaller forecast errors for the male series but not for the female series.

\subsection{Comparison with alternative estimation and forecasting methods}\label{sec:7.5}

We compared the dynamic functional principal regression method with alternative estimation and forecasting techniques and summarized the results in this section.

\subsubsection{Raw vs smoothed mortality data}

\begin{figure}[!htb]
\centering
\includegraphics[width=0.49\linewidth]{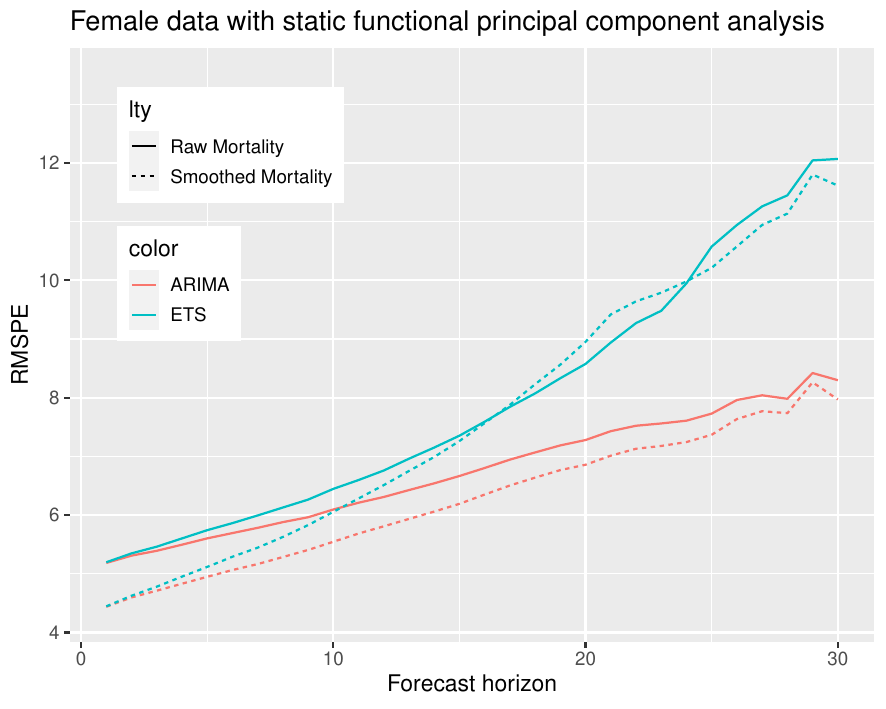}\label{fig2a}
\includegraphics[width=0.49\linewidth]{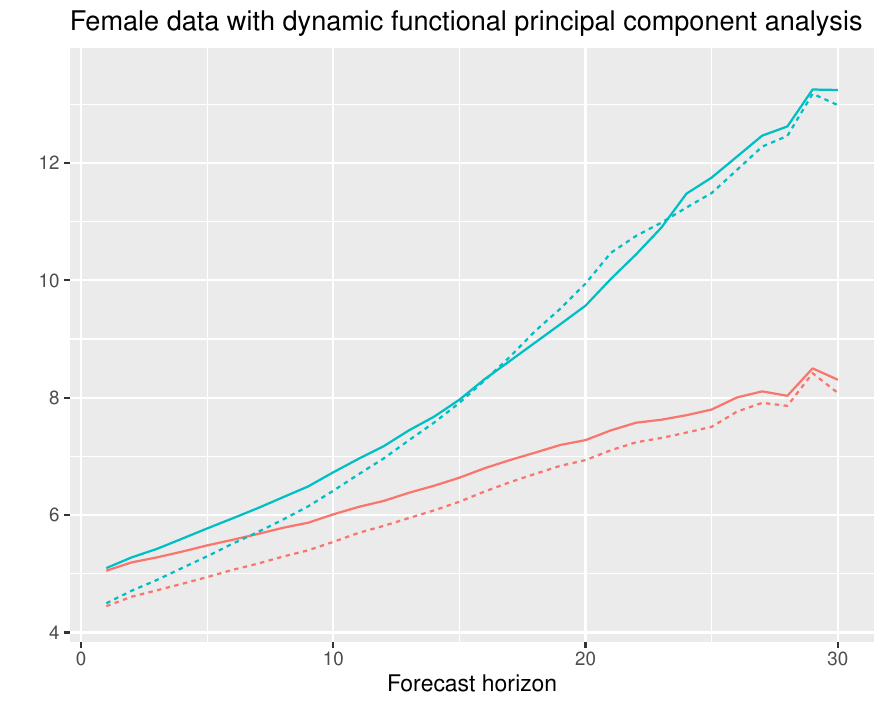}\label{fig2b} \\
\includegraphics[width=0.49\linewidth]{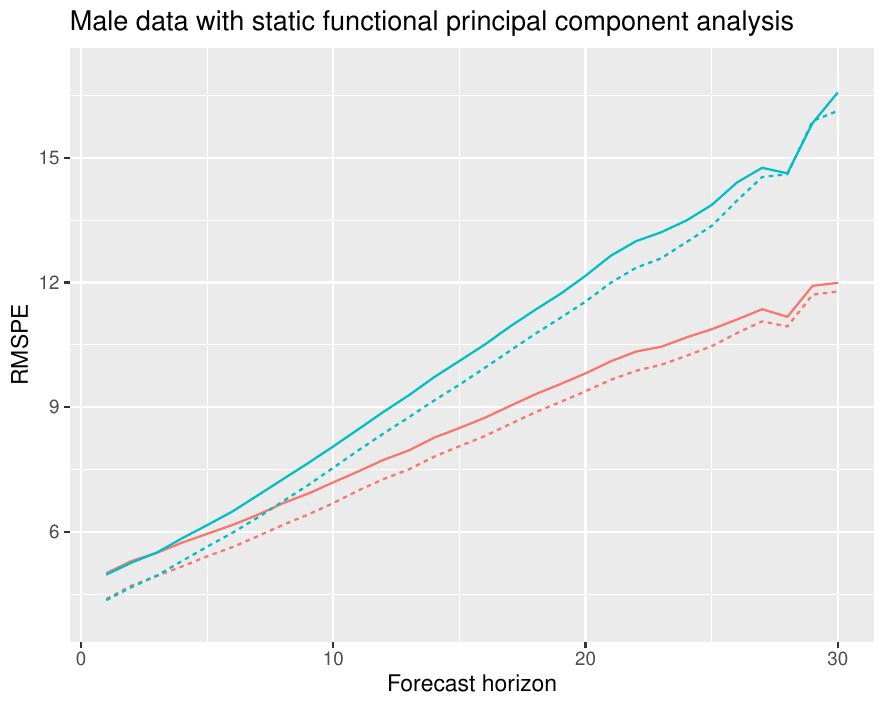}\label{fig2c} 
\includegraphics[width=0.49\linewidth]{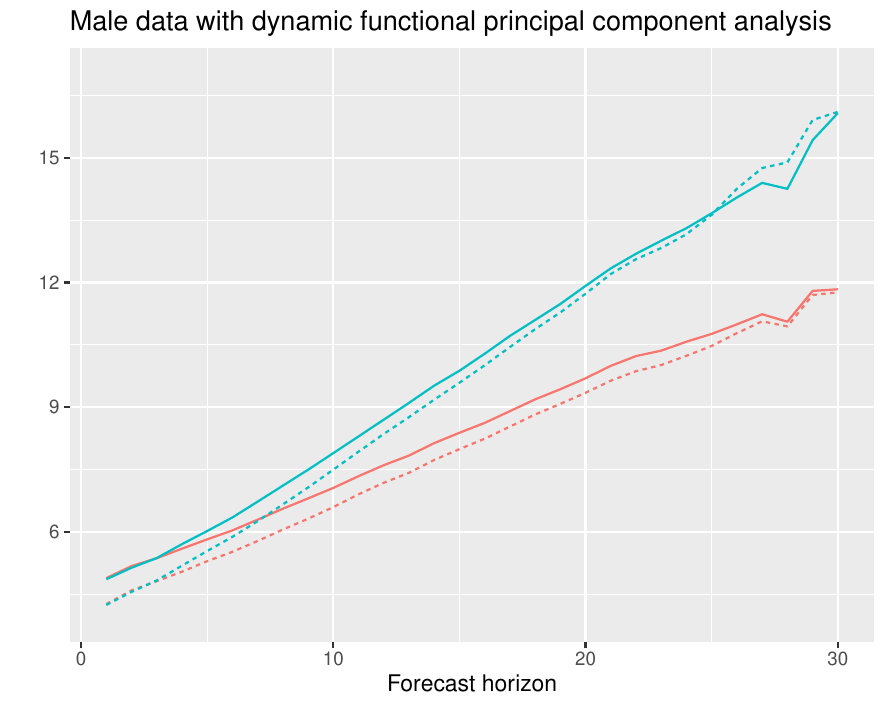}\label{fig2d}
\caption{Comparison of the point forecast accuracy between the static and dynamic functional principal component analysis with and without pre-smoothing.}\label{fig:7}
\end{figure}

First, we investigate the impact of pre-smoothing mortality rates on forecasting accuracy. Functional data analysis methods assume continuous functions in $\mathcal{L}^2(\mathcal{I})$ embedding the discrete mortality observations, which are considered contaminated by measurement errors \citep{RS06}. The nonparametric smoothing technique with monotonic constraints for ages over 65 advocated by \cite{HU07} has been widely considered in recovering the underlying smooth mortality functions from the raw observations \citep[see, e.g.,][]{shang2017grouped,shang2021forecasting}. We apply this smoothing technique to Swedish female and male mortality data before conducting the modelling and forecasting analysis and report the results in Figure~\ref{fig:7}. Pre-smoothing mortality observations help eliminate some measurement errors and reduce point forecasting errors when combined with static and dynamic functional principal component analysis methods. Interval forecasts display similar patterns, indicating the pre-smoothing step contributes to more accurate forecasts, which plots are omitted in the article to save space and can be provided upon request. We present and discuss the results obtained using the smoothed mortality data in Sections~\ref{sec:7.3} and~\ref{sec:7.4}.

\subsubsection{Static vs dynamic functional principal component analysis}

\begin{figure}[!htb]
\centering
\includegraphics[width=0.49\linewidth]{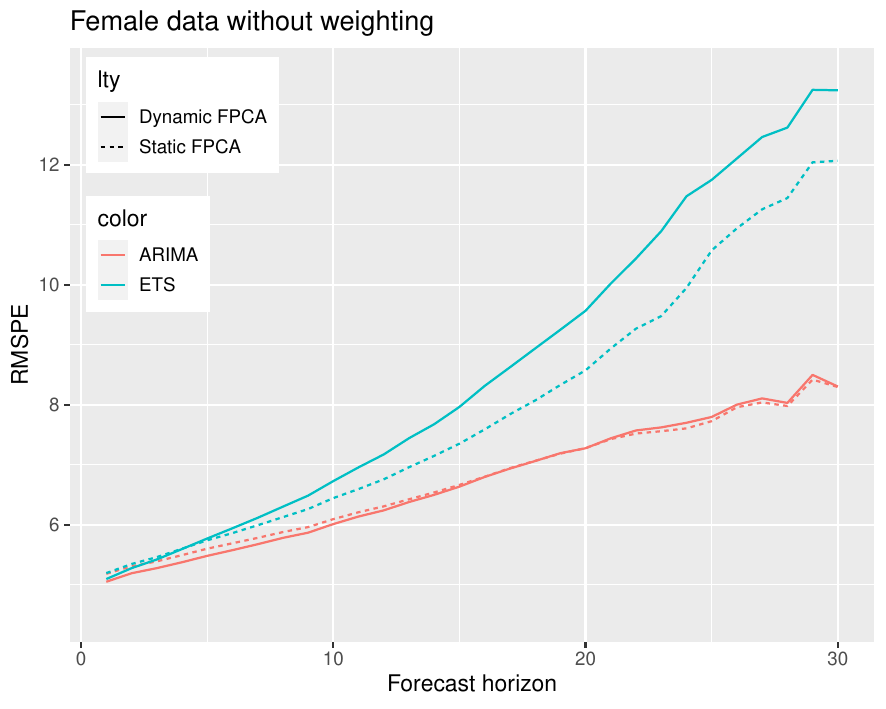}\label{fig1a}
\includegraphics[width=0.49\linewidth]{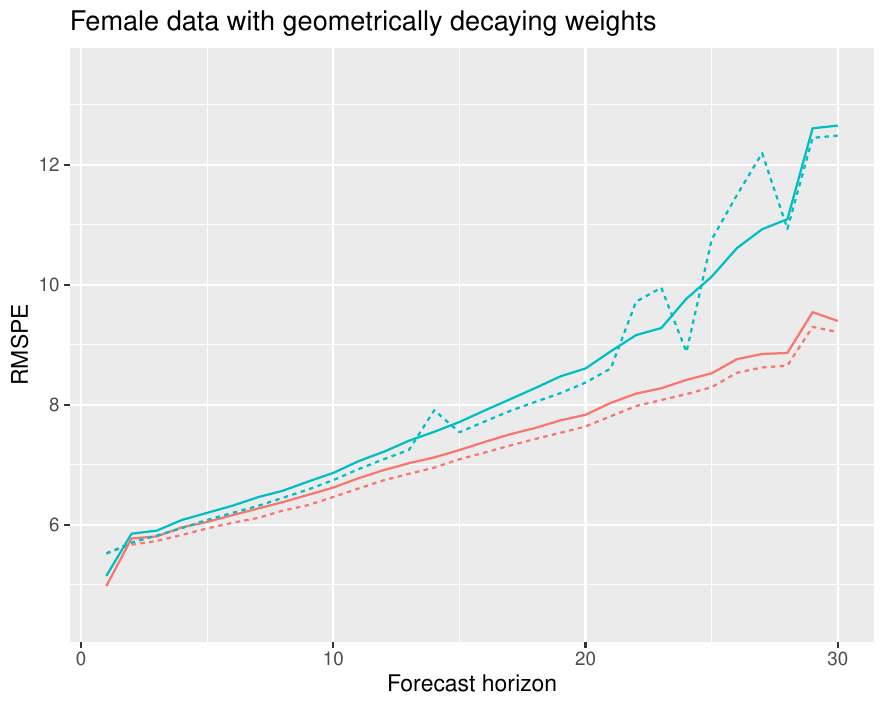}\label{fig1b} \\
\includegraphics[width=0.49\linewidth]{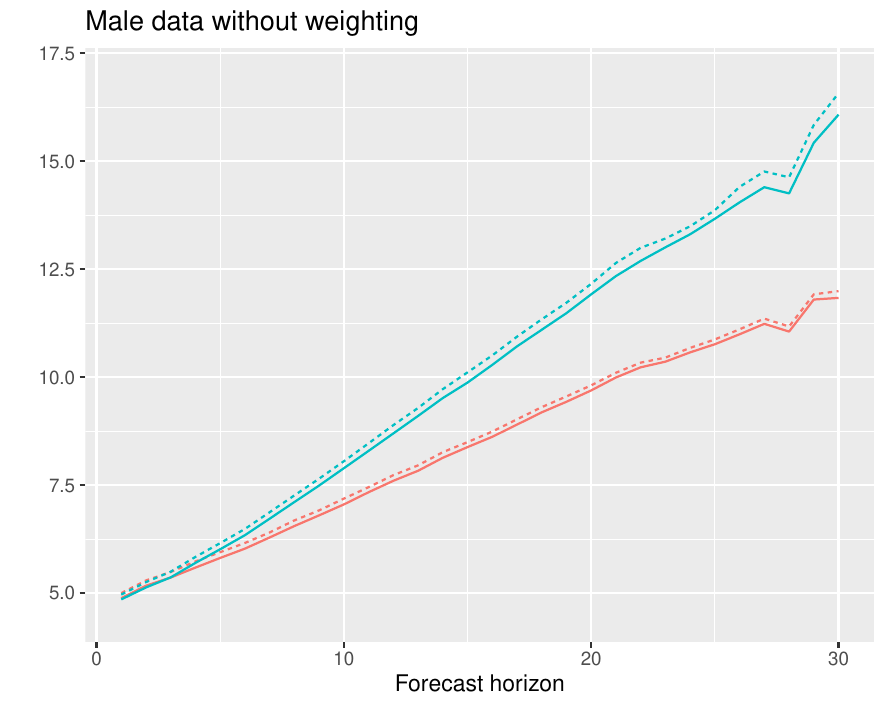}\label{fig1c} 
\includegraphics[width=0.49\linewidth]{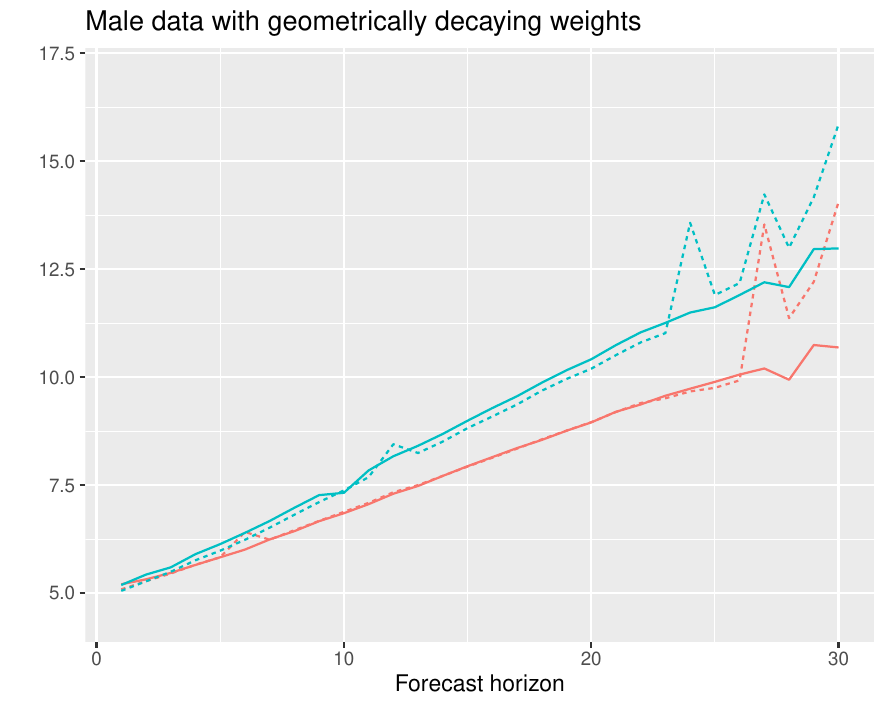}\label{fig1d}
\caption{Comparison of static and dynamic FPCA in point forecast accuracy. The static functional principal components are extracted from the sample variance function. The dynamic functional principal components are extracted from the estimated long-run covariance function based on a kernel sandwich estimator with the plug-in bandwidth.}\label{fig:8}
\end{figure}

We compare the performances of the static and dynamic functional principal component analysis methods regarding forecasting accuracy. Figures~\ref{fig:7} and~\ref{fig:8} reveal that the dynamic functional principal component analysis method contributes to more accurate male forecasts. In contrast, the static functional principal component analysis method produces more precise female forecasts. In addition, when geometrically decaying weights are used to make forecasts, dynamic functional principal component analysis can produce more stable forecasts than its static counterpart. An explanation is that for the dynamic method to work well, it is essential to accurately estimate the long-run covariance function. We choose a kernel sandwich estimator that combines variance with weighted autocovariance at data-driven lags. Since the estimated autocovariance function can be positive or negative-valued, the cancellation effect is likely to occur. Therefore, we decide to use the dynamic functional principal component analysis method in this study. Interval forecast results with similar patterns are provided upon request.

\subsubsection{Univariate vs multivariate methods for forecasting principal component scores}

\begin{figure}[!htb]
\centering
\includegraphics[width=0.49\linewidth]{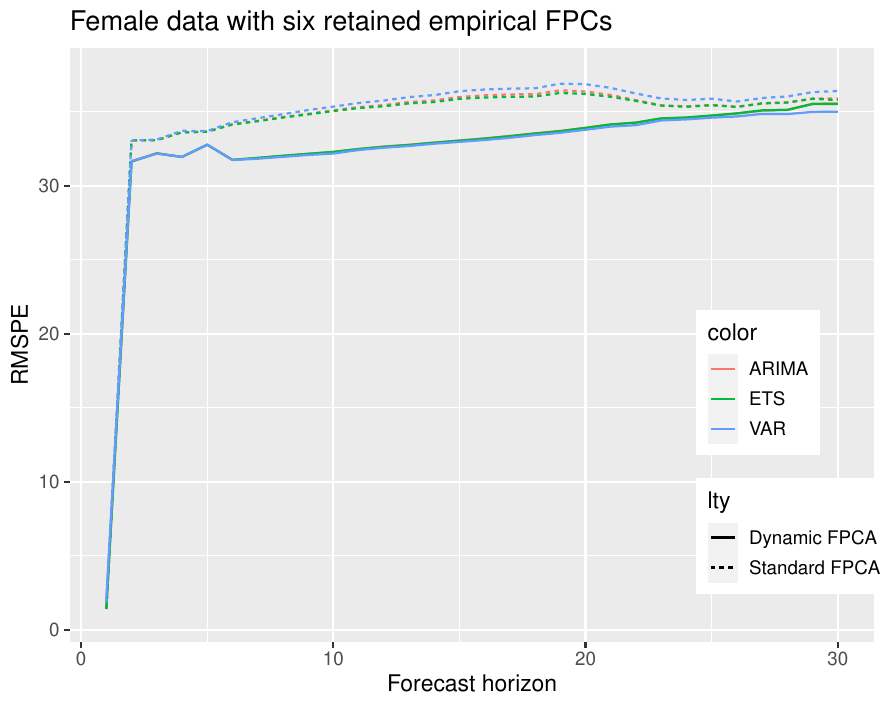}\label{fig4a}
\includegraphics[width=0.49\linewidth]{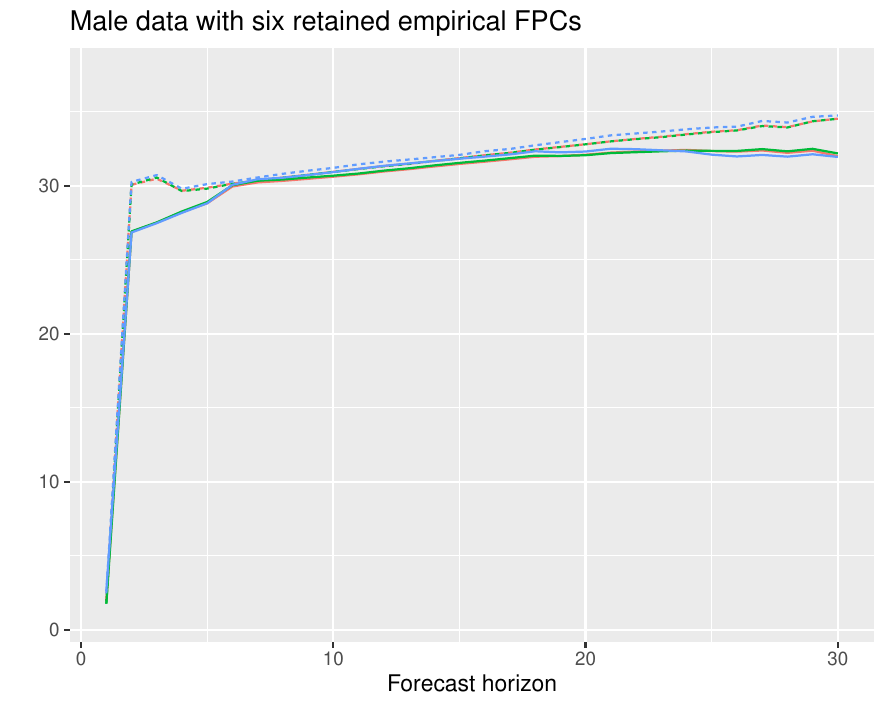}\label{fig4b} \\
\includegraphics[width=0.49\linewidth]{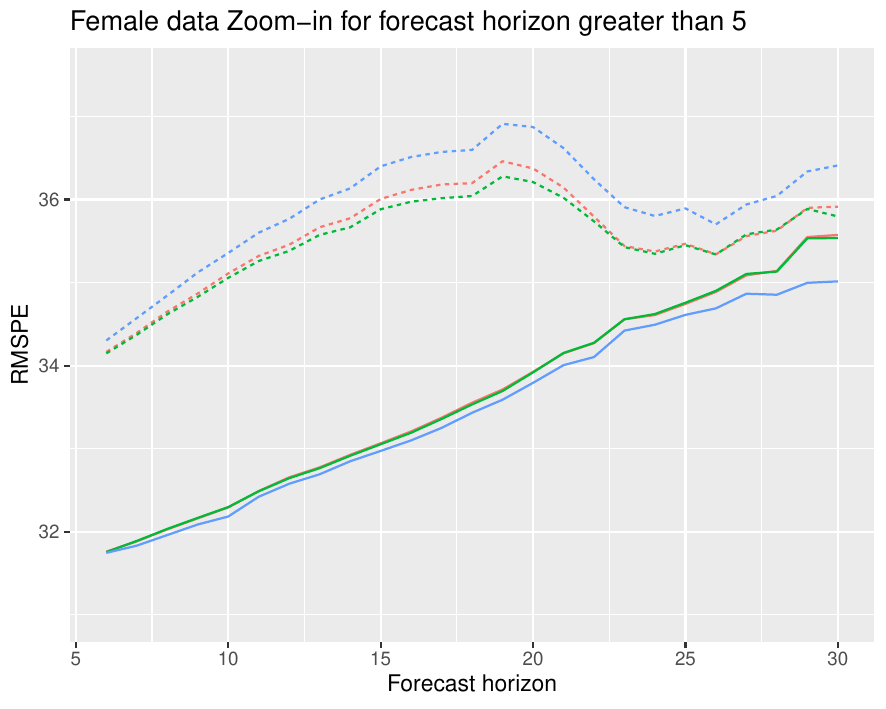}\label{fig4c}
\includegraphics[width=0.49\linewidth]{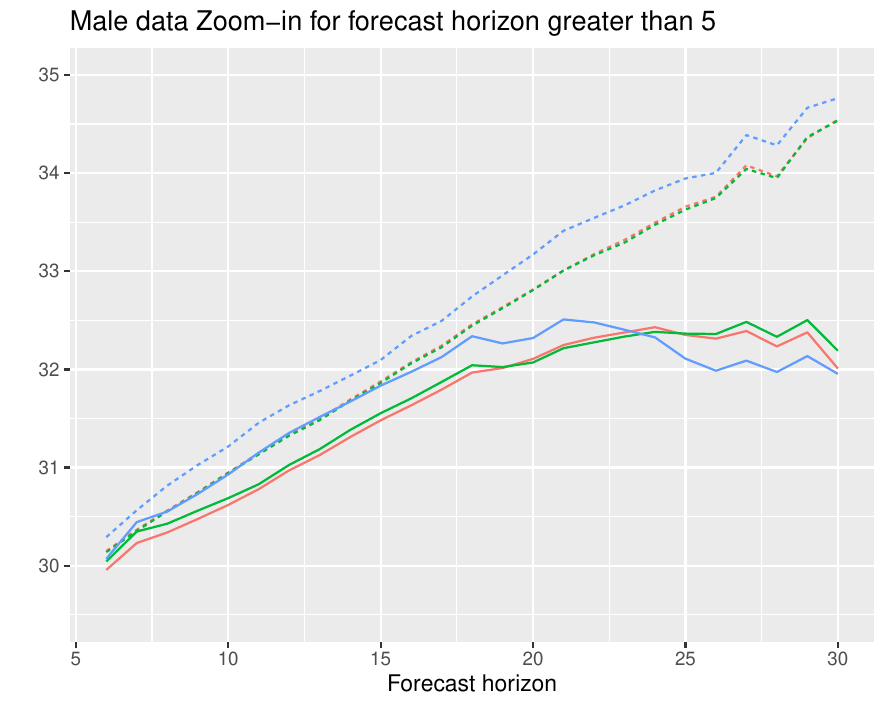}\label{fig4d}
\caption{Comparison of static and dynamic functional principal component analysis (FPCA) regarding point forecast accuracy.}\label{fig:9}
\end{figure}

The univariate time series forecasting method has been commonly used in the projection of functional principal component scores since the success of \cite{HS09}. The multivariate functional principal component method \citep{chiou2016multivariate,happ2018multivariate} as an alternative method is also used in modelling and forecasting mortality rates \citep[see, e.g.,][]{lam2023multipopulation}. Using the first six empirical functional principal components, we compare the performances of univariate time series forecasting models and vector autoregressive models in Figure~\ref{fig:9}. It can be seen that univariate forecasting methods (i.e., ARIMA and ETS) generally outperform the VAR model except for male series at forecast horizons over $h=23$. The overall differences in RMSPE between the univariate and multivariate are only marginal. Therefore, we select univariate time series models to project the dynamic functional principal components described in Section~\ref{sec:3.4}.

\subsubsection{Female vs male empirical dynamic functional principal components}

\begin{figure}[!htb]
\centering
\includegraphics[width=17.5cm]{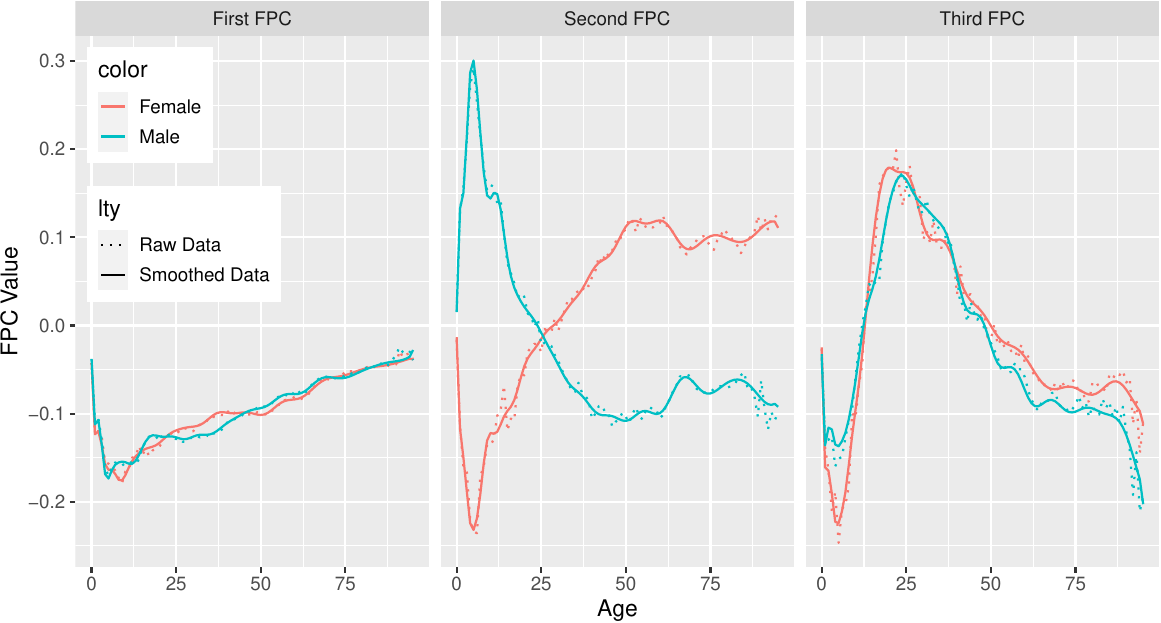}
\caption{Comparison of the first three empirical dynamic functional principal components (FPCs) obtained from the raw and smoothed age-specific mortality data from 1751 to 1993.}\label{fig:10}
\end{figure}

Finally, we compare the empirical functional principal components for the Swedish female and male mortality series. Figure~\ref{fig:10} illustrates the $k = 1, 2, 3$ empirical functional principal components obtained from the training and validation datasets (i.e., 1751--1993). The first functional principal component summarises the major differences in age-specific mortality rates between females and males because of the estimated number of components $\widehat{K} = 1$ determined by the eigenvalue ratio method \citep[see, e.g.,][]{AH13, LRS20, MGG22}. The empirical dynamic functional principal component scores are positive for the first and the third empirical functional principal components and negative for the second functional principal component.

The different shapes of FPC curves in the left-most panel of Figure~\ref{fig:9} reveal various mortality patterns for Swedish females and males. For adolescents and young adults aged 10--24 years, males generally have higher mortality rates than females. This gender gap in mortality for young Swedens is mainly attributed to higher male deaths caused by accidental injury and other avoidable causes \cite{kiadaliri21, JLAJ06}. In contrast, young and middle-aged females aged between 25 and 48 years are observed to have higher mortality than males. This disparity in mortality can be explained by the fact that Swedish females less than 55 years old are more likely to have congenital heart disease, obesity, hypertrophic cardiomyopathy, depression, and cancer than males \cite{Basic22}, contributing to higher fatalities for women. The female and male functional principal component curves gradually converge after age 60, confirming the decreasing gender gap in life expectancy in Sweden \citep{SAFF18}. The mortality gender gap in Sweden is one of Europe's lowest \citep{van2013gender}. Slight differences exist between male and female groups in the second and third empirical dynamic functional principal components. However, in this study, the first functional principal component is the only component retained in the modelling and forecasting process according to \eqref{eq:3.2}. Hence, the first functional principal component curves provide sufficient information on the mortality differences.

\section{Conclusion}\label{sec:8}

We extend the forecasting method of \cite{MGG22} by considering geometrically decaying weights for modeling and forecasting nonstationary functional time series. The method applies two-stage functional principal component decompositions to extract basis functions for nonstationary and stationary components. We obtain the $h$-step-ahead forecast curves by forecasting the projected scores and determine the point and interval forecast accuracies for $h=1,2,\dots,30$. We evaluate and compare the forecast accuracy between the standard and weighted forecasting methods. The standard method provides smaller forecast errors than the weighted one for the Swedish female data. In contrast, the weighted method helps forecast the Swedish male data, with smaller point and interval forecast errors than the standard method. This finding confirms that the weighted extension has merit in improving forecast accuracy for modelling a long curve time series. The implementation of our proposed method is available at \url{https://github.com/YangANU/Nonstationary_functional_time_series_forecasting}.

There are at least two ways in which the methodology can be extended:
\begin{enumerate}
\item[1)] Instead of down-weighting the data from the distant past, one could apply a change-point detection method of \cite{ARS18} to identify the change point before applying the method of \cite{MGG22}.
\item[2)] The idea of geometrically decaying weights can be applied to other countries and can be extended to model the age distribution of death counts. By themselves, the objects reside in a nonlinear simplex. 
\end{enumerate}

\section*{Acknowledgment}

The authors would like to thank Editor, Associate Editor, and reviewers for their constructive comments that helped improve the manuscript greatly. The first author is partially supported by the Australian Research Council Discovery Project (DP230102250) and Future Fellow (FT240100338). The usual disclaimer applies.

\bibliographystyle{agsm}
\bibliography{DFPCA_GW.bib}

\end{document}